\documentclass[11pt]{article}
\usepackage[a4paper,margin=2.5cm]{geometry}
\usepackage{amsmath}
\usepackage{amssymb}
\usepackage{graphicx}
\usepackage{lineno}
\usepackage[onehalfspacing]{setspace}
\usepackage{cite}
\usepackage[hidelinks]{hyperref}
\allowdisplaybreaks

\begin{document}

\newcommand{\Motwopi}{28.2604}
\newcommand{\Motwopiup}{28.3}
\newcommand{\abarcrit}{0.05238}
\newcommand{\ablstageone}{0.065827}
\newcommand{\ablstagethreeA}{-0.0047405}
\newcommand{\ablstagethreeB}{-0.00103975}
\newcommand{\ablstagethreeZ}{-0.00148259}
\newcommand{\ablstagetwo}{0.0033249}
\newcommand{\acabarfour}{0.40349}
\newcommand{\acabarzero}{0.40142}
\newcommand{\alphasScanA}{0.0524}
\newcommand{\alphasScanB}{0.105}
\newcommand{\alphasScanC}{0.209}
\newcommand{\alphasScanD}{0.314}
\newcommand{\archiveprinted}{123}
\newcommand{\bbarvalue}{0.69444444}
\newcommand{\bessconst}{2.7094253}
\newcommand{\cauchyrad}{0.25}
\newcommand{\cellsA}{12}
\newcommand{\cellsB}{9}
\newcommand{\cellsC}{6}
\newcommand{\cellsD}{3}
\newcommand{\cellsE}{0}
\newcommand{\cellstotal}{30}
\newcommand{\cellswideD}{5}
\newcommand{\cellswideE}{2}
\newcommand{\cellswidetotal}{26}
\newcommand{\chionepp}{-642.5264}
\newcommand{\chizeropp}{33.6576}
\newcommand{\clayeronebound}{\tfrac{12451}{4050}-\tfrac{\pi^2}{12}}
\newcommand{\clayertwoinf}{\tfrac{157}{210}}
\newcommand{\convfaileps}{0.4594}
\newcommand{\convfailratio}{-0.3395}
\newcommand{\damphi}{0.0843}
\newcommand{\damplo}{0.00234}
\newcommand{\dthirteenval}{-\tfrac12}
\newcommand{\inversionabscissahi}{110.5}
\newcommand{\inversionabscissalo}{6.91}
\newcommand{\lobeaphi}{0.000631}
\newcommand{\lobeaplo}{0.000483}
\newcommand{\lobematchhi}{0.00339}
\newcommand{\lobematchlo}{0.000514}
\newcommand{\lobevarhi}{0.0321}
\newcommand{\lobevarlo}{0.00123}
\newcommand{\maxPwideAd}{2.11921}
\newcommand{\maxPwideAdnu}{2.25}
\newcommand{\plusabarD}{10^{-4}}
\newcommand{\plusepshiA}{0.1951}
\newcommand{\plusepshiD}{0.0003866}
\newcommand{\plusepsloA}{0.4923}
\newcommand{\plusepsloD}{0.02469}
\newcommand{\plusratiohiD}{-0.3334}
\newcommand{\plusratioloD}{-0.1693}
\newcommand{\posgapA}{2.2\times10^{-6}}
\newcommand{\posgapB}{6.3\times10^{-4}}
\newcommand{\posgapC}{0.0282}
\newcommand{\posgapD}{0.158}
\newcommand{\ratiohi}{11.86}
\newcommand{\ratiolo}{427.7}
\newcommand{\ratiomatchhi}{14.12}
\newcommand{\rgiDeltanu}{0.25}
\newcommand{\rgiYa}{5}
\newcommand{\rgiYb}{2.5}
\newcommand{\rgiYc}{1.25}
\newcommand{\rgiYd}{0.833}
\newcommand{\rgieigbound}{2.4\times10^{-6}}
\newcommand{\rgieigboundwide}{1.0\times10^{-5}}
\newcommand{\rgimeAaA}{-0.00265593}
\newcommand{\rgimeAbA}{-0.0851594}
\newcommand{\rgimeAcA}{-0.519895}
\newcommand{\rgimeAdA}{-0.554882}
\newcommand{\rgimeBaA}{-0.00128987}
\newcommand{\rgimeBbA}{-0.0432302}
\newcommand{\rgimeBcA}{-0.266545}
\newcommand{\rgimeBdA}{-0.331953}
\newcommand{\rgimeZaA}{-0.00160849}
\newcommand{\rgimeZbA}{-0.0527008}
\newcommand{\rgimeZcA}{-0.320726}
\newcommand{\rgimeZdA}{-0.379638}
\newcommand{\rgiwideBdE}{-0.00625724}
\newcommand{\rgiwideZdE}{-0.000245352}
\newcommand{\rgiwitBform}{-0.00128987}
\newcommand{\rgiwitBfour}{-0.160463}
\newcommand{\rgiwitBone}{0.495854}
\newcommand{\rgiwitBthree}{0.495854}
\newcommand{\rgiwitBtwo}{-0.675841}
\newcommand{\rgiwitBzero}{-0.160463}
\newcommand{\rootcountedge}{-3}
\newcommand{\rootcountfar}{0.006}
\newcommand{\rootcountmargin}{0.01}
\newcommand{\rootcountpoints}{60}
\newcommand{\rootcountreach}{20000}
\newcommand{\rootcountreachsmall}{400}
\newcommand{\rootcountrzero}{1238.65}
\newcommand{\rootcountwidefreq}{21}
\newcommand{\rootcountwidepoints}{252}
\newcommand{\scanabarA}{0.05}
\newcommand{\scanabarB}{0.10}
\newcommand{\scanabarC}{0.20}
\newcommand{\scanabarD}{0.30}
\newcommand{\scantA}{0.25}
\newcommand{\scantB}{0.50}
\newcommand{\scantC}{1.00}
\newcommand{\scantD}{2.00}
\newcommand{\scantE}{4.00}
\newcommand{\semidefAc}{-0.921618}
\newcommand{\semidefBa}{-0.10446}
\newcommand{\shiftb}{-2.92}
\newcommand{\shiftphi}{-23.73}
\newcommand{\shiftpi}{52.15}
\newcommand{\stripfaileps}{0.403}
\newcommand{\stripfailratio}{-0.371}
\newcommand{\throne}{0.052383}
\newcommand{\thrsuppconst}{\tfrac15}
\newcommand{\thrthree}{0.013435}
\newcommand{\thrtwo}{0.023192}
\newcommand{\weightlimit}{4.35}
\newcommand{\weightsixteen}{3.92}
\newcommand{\widecells}{12}
\newcommand{\wideoverone}{9}
\newcommand{\zeroaphi}{6.04232}
\newcommand{\zeroaplo}{12.2475}
\newcommand{\zeromatchhi}{5.2948}

\title{Forward BFKL evolution at fixed coupling as a random walk, and the loss of its probability law at fixed next-to-leading order\thanks{This research is supported in part by the Council of Higher Education of Israel through the grant ``Support Program of the High Energy Physics''.}}
\author{Alex Prygarin\footnote{Corresponding author. E-mail: \texttt{alexanderp@ariel.ac.il}. ORCID: 0000-0002-1167-0654.}, Claudelle Capasia Madjuogang Sandeu and Karam Shekh Yusuf\\[6pt]
\small Department of Physics, Ariel University, Ariel 40700, Israel}
\date{}

\maketitle

\begin{abstract}
The BFKL equation describes how the density of gluons in a hadron grows with the collision energy. This evolution is often treated as a random walk in the logarithm of the gluon transverse momentum, but it has not been established that one set of positive step probabilities describes this walk in the size and the direction of the momentum, or that such probabilities exist beyond leading order. Monte Carlo sampling without weights needs such probabilities. We show that one positive step density describes the evolution in the size and the direction of the momentum with no approximation at leading order, fixed coupling and zero momentum transfer, once the gluon density is rescaled by the size of its transverse momentum and its growth is divided out. The displacements of the resulting walk over separate, equal intervals of rapidity, the logarithm of the energy in units of a reference energy, are then independent and identically distributed, which makes the walk a L\'evy process. We also show that this description fails when the series in the coupling is cut at next-to-leading order, with the reference energy and the scale of the coupling both set by the product of the transverse momenta before and after an emission. The evolution averaged over directions then has no probability law at any fixed positive coupling and rapidity, because a negative correction growing as the squared logarithm of the momentum ratio outweighs the leading term when one momentum is much larger than the other. A positive step density for the averaged walk agrees with this truncated evolution to next-to-leading order and defines a random walk for every value of the strong coupling below one, so the failure comes from the truncation. The main message is that agreement with perturbation theory does not guarantee positive step probabilities, so an evolution equation summed to all orders and meant to describe a random walk must be tested for them, and at fixed coupling this test can be made before the equation is used to generate events.

\end{abstract}

\section{Introduction}
\label{sec:intro}

The BFKL equation of Kuraev, Lipatov and Fadin \cite{KLF} and Balitsky and Lipatov \cite{BL} describes the growth of the gluon density as a power of the energy, with an exponent proportional to the coupling at leading order. Successive emissions also change the gluon transverse momentum. Bartels and Lotter \cite{BartelsLotter} and Forshaw and Ross \cite{ForshawRoss} describe the Gaussian diffusion of the leading-order Green function, the solution that starts from a single transverse momentum. This diffusion appears at large rapidity, the logarithm of the energy that plays the role of the evolution time, and supports the picture of a random walk in logarithmic transverse momentum. A random walk requires positive step probabilities. These probabilities are needed for unweighted Monte Carlo sampling and matter when a stochastic picture of linear evolution motivates the passage to saturation, where interactions at high gluon densities slow the growth. We ask whether the forward equation, with zero momentum transfer, defines a walk with positive step probabilities at fixed coupling, without a diffusion approximation.

Mueller \cite{MuellerDipole} gives a different probabilistic description in terms of a cascade of splitting color dipoles, which reproduces the same linear evolution. Probability also enters the interpretation of the nonlinear regime. The equations of Balitsky \cite{Balitsky} and Kovchegov \cite{Kovchegov} describe saturation of the scattering amplitude. Munier and Peschanski \cite{MunierPeschanski} relate Balitsky-Kovchegov evolution, after expanding the eigenvalue about the symmetric point, to traveling waves of the Fisher-Kolmogorov-Petrovsky-Piscounov equation.

Levin and Lublinsky \cite{LevinLublinsky} derive a linear equation for the dipole generating functional, and Kozlov, Levin and Prygarin \cite{KozlovLevinPrygarin} develop a probabilistic interpretation of the BFKL Pomeron calculus. Iancu, Soyez and Triantafyllopoulos \cite{IancuSoyezTriantafyllopoulos} show that, at large $N_c$, the Pomeron-loop equations cannot describe dipoles that both split and recombine with positive probabilities. They conclude that dipole recombination cannot represent saturation even effectively, and note that stochastic descriptions in other degrees of freedom are not excluded. Hatta and Iancu \cite{HattaIancu} find that the collinearly improved Balitsky-Kovchegov kernel does not factorize into two independent emissions. It lacks the corresponding Fokker-Planck Hamiltonian form and, they argue, a Langevin formulation local in rapidity. They construct a Langevin equation by giving up that locality, and they question whether the probabilistic interpretation of the Color Glass Condensate can be maintained beyond leading order.

Positivity problems beyond leading order were established by Bl\"umlein and Vogt \cite{BlumleinVogt}, Ross \cite{Ross}, Levin \cite{Levin} and Salam \cite{Salam}. Their results concern evolution solutions, amplitudes or structure functions, and identify the role of double transverse logarithms and fixed-order truncation. Levin locates the coupling where the next-to-leading diffusion coefficient changes sign \cite{Levin}. Resummations developed by Salam \cite{Salam}, Ciafaloni and Colferai \cite{CiafaloniColferai}, Ciafaloni, Colferai and Salam \cite{CCS}, Sabio Vera \cite{SabioVera}, and Colferai, Li and Sta\'sto \cite{ColferaiLiStasto} address the behavior at strongly ordered transverse momenta. In particular, the all-poles approximation of Sabio Vera removes the oscillations of the Green function produced by the fixed-order expression \cite{SabioVera}. The diffusion picture does not settle the separate question of whether one non-negative step measure generates the evolution in both the magnitude and the direction of the transverse momentum, and whether such a measure exists beyond leading order.

Other probabilistic descriptions in high-energy QCD help place this question. A L\'evy process is a random walk in continuous time whose steps over separate intervals of equal duration are independent and identically distributed. Caucal and Mehtar-Tani \cite{CaucalMehtarTani} use the name L\'evy in its other sense, for a heavy-tailed distribution of the transverse momentum produced by double-logarithmic corrections. Their result concerns the asymptotic solution of the evolution equation for transverse-momentum broadening in a medium, at fixed coupling and large system size, with a saturation scale that grows faster than diffusively. Angular-ordered cascades were formulated by Ciafaloni \cite{CiafaloniCCFM} and by Catani, Fiorani and Marchesini \cite{CFMnp}, and Jung and Salam \cite{JungSalam} implement them in a Monte Carlo generator. Giovannini and Van Hove \cite{GiovanniniVanHoveClan} use infinite divisibility to describe multiplicity distributions. An infinitely divisible law can be split into any number of independent, identically distributed pieces. This paper studies the step measure of the linear forward BFKL kernel itself, whose density falls off exponentially in the logarithm of the momentum ratio, the variable in which the walk takes its steps (Section~\ref{sec:generator}, Eq.~\eqref{eq:rho}).

Marchesini and Onofri \cite{MarchesiniOnofri} rescale the amplitude and remove its growth to obtain the gain-loss equation used here at zero conformal spin, where the conformal spin labels the angular harmonic of the walk and zero spin is the angle-averaged mode. They write it in a logarithmic impact-parameter variable in place of logarithmic transverse momentum, and its kernel is positive. Artru et al.\ \cite{ArtruEtAl} formulate BFKL evolution as a master equation in Sec.~4.3.4 of their review, reweighting the density by a momentum power and normalizing its integral. Their short-range approximation combines growth, drift and diffusion to give an intercept independent of that power. Their family excludes the rescaling by the momentum magnitude used here and has no conformal-spin decomposition, L\'evy measure or infinite divisibility. Del Duca and Schmidt \cite{DelDucaSchmidt2} sum the conformal spins and integrate over the radial transform variable to obtain the real-emission function with its full momentum-ratio and azimuthal dependence. The same function multiplies the pure functions of Del Duca, Dixon, Duhr and Pennington \cite{DDDP}. Its spin decomposition appears in the earlier work of Del Duca and Schmidt \cite{DelDucaSchmidt}. We identify this real-emission function as one L\'evy measure on the cylinder of logarithmic transverse momentum and azimuth, which gives every angular harmonic, whereas a real-emission term with a virtual subtraction does not by itself guarantee a probability law.

In this paper we construct that law and test the next-to-leading evolution at the level of its generator, the operator that gives the rate of change of the distribution with rapidity. Every positive symmetric step measure requires the BFKL eigenvalue, taken at real values of its argument between its collinear poles, to be smallest at the symmetric point. We test this condition in the symmetric scheme, where the energy scale and the scale of the coupling are both set by the product of the transverse momenta before and after an emission. A control density with a Gaussian damping factor, positive before truncation, shows that cutting its exponent at first order can by itself destroy positivity.

Section~\ref{sec:generator} derives the leading-order density and relates its jumps to diffusion and angular relaxation. Section~\ref{sec:obstruction} establishes the fixed-order obstruction and the moment identity, and constructs a positive density matching the computed order. Section~\ref{sec:removal} tests whether conjugations and the collinear resummations in use remove the obstruction, and Section~\ref{sec:discussion} discusses the physical consequences. Appendix~\ref{app:exponent} derives the characteristic exponent. Appendices~\ref{sec:supp-measure} to~\ref{app:numerics} give the further derivations, computed details and computational protocol.

The main result of this paper is that forward BFKL evolution at leading order and fixed coupling is a L\'evy process with no approximation, once the gluon density is rescaled by the transverse-momentum magnitude and its growth is removed. One positive step measure, Eq.~\eqref{eq:rho}, describes every angular harmonic and gives the intercept, the diffusion and the azimuthal relaxation, so the random-walk picture of leading-order evolution rests on a probability law. We also show that the angle-averaged evolution truncated at next-to-leading order in the symmetric scheme has no probability law at any fixed positive coupling and rapidity. The cause is a negative double logarithm in the next-to-leading kernel, the square of the logarithm of the ratio of the transverse momenta before and after an emission. This term removes emission probability faster than the leading term produces it when one momentum is much larger than the other. As a result, the eigenvalue of the kernel violates the condition \eqref{eq:stripineq} that every positive step measure imposes. The coefficient of this double logarithm does not depend on the numbers of colors and quark flavors, so the obstruction holds whatever these numbers are. A positive density for the angle-averaged walk agrees with the truncated density to that order and is a L\'evy density for every QCD coupling $\alpha_s<1$, so the obstruction comes from the truncation. The main message is that matching the perturbative eigenvalue does not guarantee positive step probabilities, so a resummed kernel meant to describe a random walk must be tested for them, and at fixed coupling this test can be made before the kernel is used to generate events in a Monte Carlo program.

\section{The leading-order kernel as a generator}
\label{sec:generator}

Each real emission is a step that multiplies the magnitude of the gluon transverse momentum and turns its direction, the reggeization term removes weight from the starting momentum, and the excess of gain over loss is the growth that is divided out later. The forward BFKL equation at fixed coupling \cite{KLF,BL} evolves the gluon density according to
\begin{equation}
\frac{\partial f(\mathbf k,Y)}{\partial Y}
=\bar\alpha\int\frac{d^2\mathbf q}{\pi\,(\mathbf k-\mathbf q)^2}
\Big[f(\mathbf q,Y)-\frac{\mathbf k^2}{\mathbf q^2+(\mathbf k-\mathbf q)^2}
f(\mathbf k,Y)\Big] ,
\label{eq:bfkl}
\end{equation}
where $\bar\alpha$ is the coupling scaled as in \eqref{eq:conventions} below, the first term inside the bracket describes real emission of a gluon with transverse momentum $\mathbf k-\mathbf q$, and the second accounts for reggeization of the exchanged gluon.

We denote the momenta before and after an emission by $\mathbf k$ and $\mathbf k'$ and define the step variables by
\begin{equation}
\delta=\ln\frac{k'^2}{k^2},\qquad \phi=\theta_{k'}-\theta_k,\qquad
a_n=\frac{1+|n|}{2},\qquad \bar\alpha=\frac{\alpha_sN_c}{\pi} ,
\label{eq:conventions}
\end{equation}
where $\delta$ measures the change in logarithmic transverse momentum and $\phi$ the change in azimuth. Their pair $(\delta,\phi)$ lies on the cylinder
\begin{equation}
\mathbb G=\mathbb R\times S^1 ,
\label{eq:group}
\end{equation}
where multiplication of momentum magnitudes becomes addition in $\delta$. Addition of steps therefore makes $\mathbb G$ a locally compact abelian group, with dual variables $\nu\in\mathbb R$ and $n\in\mathbb Z$, the variables of the transform. Haar measure on $\mathbb G=\mathbb R\times S^1$, the flat measure invariant under translations, is $d\ell\,d\theta$, where $\ell=\ln k^2$. The momentum-space Jacobian is
\begin{equation}
d^2\mathbf k=\tfrac12k^2\,d\ell\,d\theta ,
\label{eq:jacobian}
\end{equation}
and the conjugation by the momentum magnitude given below symmetrizes the real-emission term including this Jacobian.

\subsection*{Infinite divisibility}

A L\'evy process has stationary, independent increments in continuous time. Rapidity plays the role of time here. The law at $Y$ can be subdivided if the evolution from zero to $Y$ is a L\'evy process. For every integer $N$, it can then be obtained by adding $N$ independent, identically distributed contributions, one from each interval of length $Y/N$. A law that permits this subdivision for every $N$ is infinitely divisible.

The requirement excludes nearly every distribution, because the subdivision must be possible for every $N$ and not for one $N$ alone. A Gaussian satisfies it by assigning variance $1/N$ of the whole to each piece, and a Poisson law does so with rate $1/N$ of the whole. A distribution on $\mathbb R^d$ confined to a bounded set fails unless it is concentrated at a point. A compact group factor changes this conclusion, because its Haar measure is its own convolution root for every $N$.

The L\'evy-Khintchine theorem expresses infinite divisibility through the characteristic function. An infinitely divisible law whose characteristic function never vanishes is the law of a sum of a constant drift, a Gaussian part and independent jumps, whose rate for each jump size is given by one non-negative measure. More precisely, a law with a characteristic function that never vanishes is infinitely divisible if and only if that function is $e^{-\Psi}$, where
\begin{equation}
\Psi(\xi)=q_0+i\langle b,\xi\rangle+Q(\xi)
+\int\Big(1-e^{-i\langle\xi,u\rangle}-i\langle\xi,h(u)\rangle\Big)\,\Pi(du) ,
\label{eq:LK}
\end{equation}
where the drift is $b$ and the non-negative quadratic form $Q$ describes the Gaussian part. The step measure $\Pi$ is non-negative and integrable against $\min(1,|u|^2)$, so it gives a finite rate outside every neighborhood of the origin. The truncation function $h$ enters a subtraction that makes the small-jump sum converge, and $h$ is needed only when the jumps cannot be summed directly. The killing rate $q_0\ge0$ vanishes for a probability law. The quadruple $(q_0,b,Q,\Pi)$ contains the L\'evy-Khintchine triplet and the killing rate. It is unique for a given $\Psi$ once $h$ is fixed, so one non-negative measure, together with the other three entries, determines the evolution at every rapidity.

The original kernel needs two changes before it can satisfy \eqref{eq:LK}. A step measure $\Pi$ must give a finite rate outside every neighborhood of the identity. The emission rate in \eqref{eq:bfkl} is infinite there, because the weight per unit $\delta$ tends to a constant as $q^2/k^2$ grows. The law must also be normalized, whereas BFKL evolution makes its total weight grow with rapidity. The conjugation of the next subsection makes the emission rate finite outside every neighborhood of the identity. Removing the intercept normalizes the law and turns the weighted reggeization loss into the plain loss of the generator given below. The process obtained after both changes has infinitely many short steps in every interval of rapidity, whereas the clan decomposition of multiplicity distributions by Giovannini and Van Hove \cite{GiovanniniVanHove,GiovanniniVanHoveClan} has finitely many clans (Appendix~\ref{sec:supp-measure}).

\subsection*{The conjugation that makes the step measure symmetric}

The conjugation balances the weight of a step and its reverse. In the variables \eqref{eq:conventions}, the flat measure $d^2\mathbf q$ contains a factor $q^2$, while the emission factor $1/(\mathbf k-\mathbf q)^2$ in \eqref{eq:bfkl} is homogeneous of degree $-2$. Their product has degree zero, making the emission term a convolution on $\mathbb G$. Its density against Haar measure is $\tfrac{q^2}{2\pi\,(\mathbf k-\mathbf q)^2}$. This density depends only on the step but is not even in $\delta$, and reverse steps have weights in the ratio $q^2/k^2$. We conjugate by the momentum magnitude $k$, with $\ell=\ln k^2$ as above,
\begin{equation}
g(\ell,\theta)=k\,f(\mathbf k) ,
\label{eq:conjugation}
\end{equation}
which replaces $q^2$ by $k\,q$. The resulting density, displayed below, is even in $\delta$ and decays at both ends, giving a symmetric measure with finite rate outside every neighborhood of the identity.

The remaining growth reflects the increasing gluon number. The conjugated Green function has total weight $e^{\omega_{\mathbb P}Y}$, where $\omega_{\mathbb P}$ is the Pomeron intercept. Dividing by this factor keeps its momentum distribution and removes the growing total. Define $\mathcal G(\delta,\phi;Y)\equiv\tfrac12\,k\,k'\,G(\mathbf k,\mathbf k';Y)$, the conjugated Green function.
Its transform is
$\hat{\mathcal G}(\nu,n;Y)=\exp[\bar\alpha Y\chi(n,\tfrac12+i\nu)]$,
with eigenvalue
\begin{equation}
\chi(n,\gamma)=2\psi(1)-\psi\Big(\gamma+\frac{|n|}{2}\Big)
-\psi\Big(1-\gamma+\frac{|n|}{2}\Big),
\label{eq:chi}
\end{equation}
and the $(0,0)$ mode sums over momentum and azimuth, giving the total growth of the conjugated Green function,
\begin{equation}
\int_{\mathbb G}\mathcal G(\delta,\phi;Y)\,d\delta\,d\phi
=e^{\bar\alpha Y\chi(0,\frac12)}=e^{4\bar\alpha Y\ln2},
\qquad \omega_{\mathbb P}=4\bar\alpha\ln2 .
\label{eq:growth}
\end{equation}
Removal of the growth \eqref{eq:growth} defines the normalized measure, with unit total weight at every rapidity,
\begin{equation}
\mu_t(d\delta,d\phi)\equiv e^{-\omega_{\mathbb P}Y}\,
\mathcal G(\delta,\phi;Y)\,d\delta\,d\phi\Big|_{t=\bar\alpha Y} ,
\label{eq:mut}
\end{equation}
and restoring the growth and undoing the momentum rescaling recovers the physical Green function through
\begin{equation}
G(\mathbf k,\mathbf k';Y)=\frac{2}{kk'}\,e^{\omega_{\mathbb P}Y}\,
\mu_{\bar\alpha Y}(\delta,\phi) .
\label{eq:greenrelation}
\end{equation}
Forward leading-order evolution at fixed coupling thus consists of the scalar growth \eqref{eq:growth} and probability transport of $g=kf$ on the cylinder.

\subsection*{The L\'evy density}

The conjugation \eqref{eq:conjugation} gives the real-emission term the density on the cylinder
\begin{equation}
\rho(\delta,\phi)=\frac{1}{4\pi\,\big[\cosh\frac{\delta}{2}-\cos\phi\big]}
=\frac{1}{2\pi}\frac{kk'}{(\mathbf k-\mathbf k')^2} ,
\label{eq:rho}
\end{equation}
which the removal of the growth \eqref{eq:growth} leaves unchanged. This density is the conjugated one-gluon emission probability and is therefore non-negative.

Del Duca and Schmidt \cite{DelDucaSchmidt2} and Del Duca, Dixon, Duhr and Pennington \cite{DDDP} give the same real-emission function (Appendix~\ref{sec:supp-measure} gives the correspondence of variables). We derive the density from the momentum-space kernel. The real-emission factor in \eqref{eq:bfkl} is $\tfrac{1}{\pi\,(\mathbf k-\mathbf q)^2}$ against $d^2\mathbf q$. The Jacobian \eqref{eq:jacobian} contributes $\tfrac12 q^2\,d\ln q^2\,d\theta_q$, and conjugation by \eqref{eq:conjugation} contributes $k/q$. Their product, for the kernel acting on $g$, is $\tfrac{kq}{2\pi\,(\mathbf k-\mathbf q)^2}$, or $\rho$ in \eqref{eq:rho}. The momentum dependence $kq/(\mathbf k-\mathbf q)^2$ is a function of the step alone, as follows from
\begin{equation}
\frac{(\mathbf k-\mathbf k')^2}{kk'}
=\frac{k}{k'}+\frac{k'}{k}-2\cos\phi
=2\Big[\cosh\frac{\delta}{2}-\cos\phi\Big] ,
\label{eq:stepfunction}
\end{equation}
which gives \eqref{eq:rho} and the cylinder on which the steps add.

Short steps determine the activity of the process. Near coincident momenta, $\rho$ has the planar form
\begin{equation}
\rho(\delta,\phi)=\frac{1}{2\pi|u|^2}
\Big[1-\frac{\delta^2}{48}+\frac{\phi^2}{12}+O(|u|^4)\Big],
\qquad |u|^2\equiv\frac{\delta^2}{4}+\phi^2 ,
\label{eq:nearid}
\end{equation}
whose singularity in \eqref{eq:rho} makes the total step rate infinite. Long radial steps are suppressed by the exponential decay $e^{-|\delta|/2}/2\pi$.

A radial resolution cut counts only steps with $|\delta|>\eta_\delta$. Their rate follows by integrating $\pi_0(\delta)=\int_{-\pi}^{\pi}\rho(\delta,\phi)\,d\phi$. The band $|\delta|\le\eta_\delta$ contains a neighborhood of the identity, and compactness of the azimuth makes a separate cut in $\phi$ unnecessary. The resolved rate is
\begin{equation}
\int_{|\delta|>\eta_\delta}\pi_0(\delta)\,d\delta=-2\ln\tanh\frac{\eta_\delta}{4}
=2\ln\frac{4}{\eta_\delta}+\frac{\eta_\delta^2}{24}+O(\eta_\delta^4),
\label{eq:activity}
\end{equation}
which diverges logarithmically as the cut is removed. The mean accumulated radial and azimuthal distances per unit $t$, however, remain finite,
\begin{equation}
\int_{\mathbb G}|\delta|\,\rho\,du=4\int_0^\infty\frac{\tau\,d\tau}{\sinh\tau}=\pi^2,
\qquad
\int_{\mathbb G}|\phi|\,\rho\,du=\frac{14\zeta_3}{\pi} .
\label{eq:moments}
\end{equation}
Equations \eqref{eq:activity} and \eqref{eq:moments} give infinite activity and finite variation. Infinitely many steps are short, and the sum of their lengths converges. The jumps can therefore be summed directly, without a compensator, the small-jump subtraction introduced through the truncation function. If used, the truncation function can be chosen odd under inversion, giving zero drift.

\begin{figure}[!ht]
\centering
\includegraphics[width=0.85\textwidth]{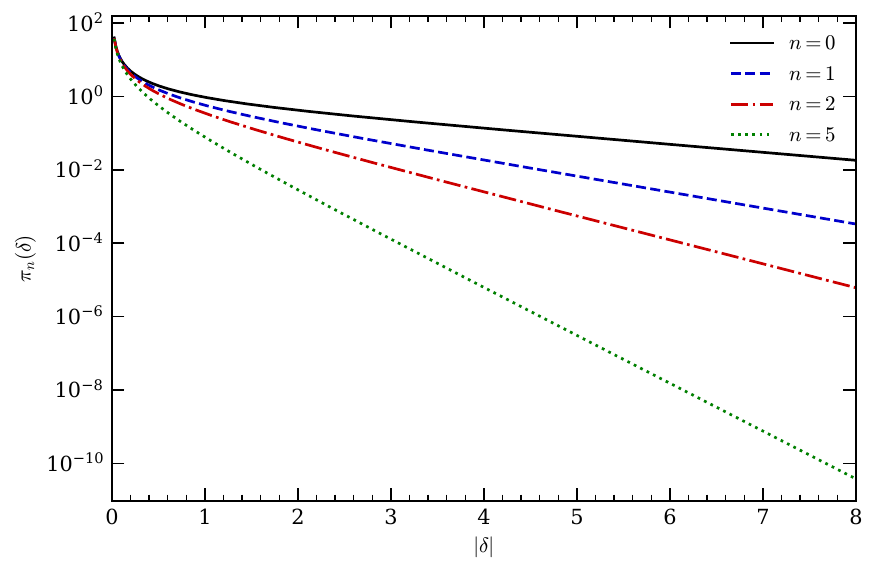}
\caption{The radial projections $\pi_n(\delta)=e^{-a_n|\delta|}/(1-e^{-|\delta|})$
of \eqref{eq:pin} at conformal spins $n=0$, $1$, $2$ and $5$, on
$0.025\le|\delta|\le8$ with a logarithmic vertical axis. The curves are ordered
by $a_n=(1+|n|)/2$ and share the $1/|\delta|$ behavior at the identity. The
identity lies outside the range over which the curves are drawn, and that
divergence is the source of the infinite activity.}
\label{fig:projections}
\end{figure}

At any finite resolution, nearly all steps are too short to see. The number of resolved steps grows as $2\ln(4/\eta_\delta)$ when the radial cut $\eta_\delta$ decreases, by \eqref{eq:activity}, but the radial steps shrink at least as fast. Their total radial variation therefore tends to $\pi^2$ per unit $t$, and the azimuthal variation is finite as well. A Gaussian component would accumulate unbounded length from arbitrarily small increments. The L\'evy-Khintchine triplet displayed below has no such component, so large-rapidity diffusion comes from the jumps. Figure~\ref{fig:projections} shows the radial projections at four conformal spins.

The measure connects to the beta family of Kuznetsov \cite{Kuznetsov}, the subordinator of M\"ohle \cite{Mohle}, the log-beta process of Walker and Muliere \cite{WalkerMuliere}, and the conformal fractional Laplacian studied by DelaTorre, del Pino, Gonz\'alez and Wei \cite{DelaTorreDelPinoGonzalezWei}, as detailed in Appendix~\ref{sec:supp-measure}.

\subsection*{The exponent at every conformal spin}

The angular Fourier coefficients of \eqref{eq:rho} resolve the density into exponential layers,
\begin{equation}
\pi_n(\delta)=\int_{-\pi}^{\pi}e^{-in\phi}\rho(\delta,\phi)\,d\phi
=\frac{e^{-|n||\delta|/2}}{2\sinh(|\delta|/2)}
=\frac{e^{-a_n|\delta|}}{1-e^{-|\delta|}}
=\sum_{m\ge0}e^{-(m+a_n)|\delta|} ,
\label{eq:pin}
\end{equation}
where the decay rates $r_{m,n}=m+a_n$ equal the distances of the collinear and anticollinear poles of $\chi(n,\gamma)$ from $\gamma=\tfrac12$. Appendix~\ref{app:exponent} derives this decomposition from the Poisson kernel. Marchesini and Onofri \cite{MarchesiniOnofri} perform the same conjugation and growth removal at zero conformal spin in an impact-parameter variable, and their gain-loss kernel is \eqref{eq:pin} at $n=0$ (Appendix~\ref{sec:supp-measure}).

Each layer with decay rate $r_{m,n}$ has finite total intensity and defines an independent compound Poisson process. Integrating \eqref{eq:rho} against $1-e^{-i(\nu\delta+n\phi)}$ using \eqref{eq:pin} gives the characteristic exponent,
\begin{equation}
q(\nu,n)=\int_{\mathbb G\setminus\{0\}}\rho(u)\big[1-e^{-i\langle\xi,u\rangle}\big]\,du
=\psi(a_n+i\nu)+\psi(a_n-i\nu)-2\psi(\tfrac12)
=\chi(0,\tfrac12)-\chi(n,\tfrac12+i\nu) .
\label{eq:q}
\end{equation}
The $n=0$ value of \eqref{eq:q} is the dispersion relation $4\ln2-\chi$ in Eq.~(3.3) of Marchesini and Onofri \cite{MarchesiniOnofri}, with their momentum variable identified as $\nu$. Appendix~\ref{app:exponent} evaluates the transform layer by layer and the zero-frequency gap between conformal spins. The positive density \eqref{eq:rho} therefore reproduces the BFKL eigenvalue in \eqref{eq:q} after intercept subtraction and sign reversal. Its layer representation is
\begin{equation}
q(\nu,n)-\kappa_n=2\sum_{m\ge0}\frac{\nu^2}{r_{m,n}\,(r_{m,n}^2+\nu^2)} ,
\qquad
\kappa_n=q(0,n)=2\big[\psi(a_n)-\psi(\tfrac12)\big] ,
\label{eq:cbf}
\end{equation}
so every term contributing to $q$ is non-negative. Appendix~\ref{app:exponent} also proves negative definiteness layer by layer, which ensures infinite divisibility.

The normalized measure $\mu_t$ of \eqref{eq:mut} is consequently an infinitely divisible probability measure on $\mathbb G$ for every $t=\bar\alpha Y$. Its L\'evy-Khintchine triplet and killing rate are
\begin{equation}
q_0=0,\qquad Q=0,\qquad b=0,\qquad
\Pi(d\delta,d\phi)=\rho(\delta,\phi)\,d\delta\,d\phi ,
\label{eq:triple}
\end{equation}
so the process is pure jump and symmetric under step inversion. The generator of the process \eqref{eq:triple} sums gains from each step and subtracts the corresponding loss,
\begin{equation}
\mathcal Af(\delta,\phi)=\int_{\mathbb G\setminus\{0\}}
\Big[f\big((\delta,\phi)+u\big)-f(\delta,\phi)\Big]\,\Pi(du) ,
\label{eq:generator}
\end{equation}
which is the form given by Hunt \cite{Hunt}, also stated by Liao \cite{Liao}, with zero drift and no Gaussian part, on a locally compact abelian group, the class of groups treated by Berg and Forst \cite{BergForst}. The finite first moments \eqref{eq:moments} make the integral absolutely convergent for bounded Lipschitz $f$. The transform on $\mathbb G$ and its dual $\hat{\mathbb G}=\mathbb R\times\mathbb Z$ is defined by
\begin{equation}
\hat\mu(\xi)=\int_{\mathbb G}e^{-i\langle\xi,u\rangle}\mu(du),
\qquad \hat\mu_t(\xi)=e^{-t\,q(\xi)},
\qquad \xi=(\nu,n),\quad t=\bar\alpha Y .
\label{eq:transform}
\end{equation}

The joint regularization of the real and virtual terms produces the subtraction in \eqref{eq:q}. The real term of \eqref{eq:bfkl} acts by convolution and the virtual term by multiplication. Each diverges separately, so they must be regulated together. At $z=(\ell,\theta)$ on $\mathbb G$, their conjugated sum is
\begin{equation}
\begin{gathered}
k\,(\mathcal Kf)(\mathbf k)=\bar\alpha\int_{\mathbb G} du\;\rho(u)
\big[g(z+u)-w(u)\,g(z)\big],\\[2pt]
w(\delta,\phi)=\frac{k\,k'}{k'^2+(\mathbf k-\mathbf k')^2}
=\frac{1}{e^{\delta/2}+2\big[\cosh\frac\delta2-\cos\phi\big]} ,
\end{gathered}
\label{eq:regulated}
\end{equation}
where the reggeization weight $w$ differs from the unit weight of a probability-conserving loss. Their integrated difference is
\begin{equation}
\int_{\mathbb G}\rho(u)\big[1-w(u)\big]\,du=4\ln2=\chi(0,\tfrac12)
\label{eq:R1}
\end{equation}
and \eqref{eq:regulated} therefore equals $\bar\alpha$ times \eqref{eq:generator} plus $\omega_{\mathbb P}$ times the identity operator. The normalization \eqref{eq:mut} removes the multiple of the identity operator, leaving $\bar\alpha$ times the generator \eqref{eq:generator}.

The central limit theorem gives the Gaussian large-rapidity limit of Marchesini and Onofri \cite{MarchesiniOnofri}, whose Eq.~(3.5) has $D=28\zeta_3$. The diffusion coefficient is the second moment of the step measure. Writing $\chi_0(\gamma)$ for the zero-spin eigenvalue $\chi(0,\gamma)$ of \eqref{eq:chi}, we find
\begin{equation}
\int_{-\infty}^{\infty}\delta^2\,\pi_0(\delta)\,d\delta
=4\sum_{m\ge0}\frac{1}{(m+\frac12)^3}=28\zeta_3=\chi_0''(\tfrac12) ,
\label{eq:secondmoment}
\end{equation}
an identity that Section~\ref{sec:obstruction} uses again as the local form of the obstruction. The Gaussian describes $\mu_t$ on the width scale $\sqrt{28\zeta_3\bar\alpha Y}$. At large $|\delta|$, the exponential step tails instead give $\mu_t$ a decay slower than any Gaussian. Appendix~\ref{sec:supp-regularity} derives the regularity of the law at small and large rapidity.

\subsection*{The intercept as a relaxation rate, and the iterative solution}

The radial size of a step controls how much it turns. For radial step $\delta$, let $r=e^{-|\delta|/2}$, the parameter $z$ in \eqref{eq:poissonkernel}. The identities $\cosh(\delta/2)=(r+r^{-1})/2$ and $\sinh(|\delta|/2)=(r^{-1}-r)/2$ give the angular distribution by dividing \eqref{eq:rho} by its $n=0$ projection in \eqref{eq:pin},
\begin{equation}
\frac{\rho(\delta,\phi)}{\pi_0(\delta)}
=\frac{\sinh(|\delta|/2)}{2\pi\,\big[\cosh(\delta/2)-\cos\phi\big]}
=\frac{1}{2\pi}\,\frac{1-r^2}{1+r^2-2r\cos\phi} ,
\label{eq:wrappedcauchy}
\end{equation}
the wrapped Cauchy density with parameter $r$, a Cauchy law on the line folded around the circle. The parameter $r=k_</k_>$ is the ratio of the smaller of $k$ and $k'$ to the larger. A step that changes the transverse-momentum magnitude by a large factor is therefore almost isotropic in azimuth, while a step that barely changes it is almost forward. The Fourier coefficients $\widehat{\rm WC}(r)_n=r^{|n|}$ show that convolutions remain in the wrapped Cauchy family.

The azimuth at rapidity $Y$, conditioned on the radial path, is therefore wrapped Cauchy with parameter $R=e^{-V/2}$. Here $V$ is the accumulated absolute radial displacement. This nondecreasing process is a subordinator with L\'evy measure $2\pi_0(v)\,dv$ on $v>0$. The relation $e^{-|n|v/2}\pi_0(v)=\pi_n(v)$ gives its Laplace exponent at $\lambda=|n|/2$ as
$\int_0^\infty(1-e^{-|n|v/2})\,2\pi_0(v)\,dv=\kappa_n$,
by \eqref{eq:kappaderiv}. Averaging over the radial path yields the angular relaxation,
\begin{equation}
\big\langle e^{in\phi}\big\rangle_t=\mathbb E\big[R^{|n|}\big]
=e^{-t\kappa_n},\qquad
\kappa_1=4\ln2,\qquad
\bar\alpha\kappa_1=\omega_{\mathbb P} .
\label{eq:kappa}
\end{equation}
The first angular harmonic of the conjugated Green function neither grows nor decays with rapidity because $\chi(1,\tfrac12)=0$. Dividing out the overall growth therefore makes the normalized harmonic in \eqref{eq:kappa} decay at the intercept per unit rapidity. Higher harmonics have rates $\kappa_n$, the zero-frequency gaps of the exponent. The same rates govern large-rapidity azimuthal decorrelation in the leading-order observable of Mueller and Navelet \cite{MuellerNavelet}, studied by Del Duca and Schmidt \cite{DelDucaSchmidt} and Sabio Vera \cite{SabioVeraMN}. They are implicit in those decorrelation formulas through the same general-spin digamma eigenvalue, and Stirling \cite{Stirling} gives them explicitly in the large-rapidity limit. Averaging over the radial path in \eqref{eq:kappa} yields them without a conformal-spin sum and gives them a direct relaxation interpretation.

A radial cut splits the process, by the L\'evy-It\^o theorem, into resolved jumps and an independent remainder of infinitely many shorter steps. This decomposition has the same shape as that of the iterative solutions of \eqref{eq:bfkl} by Schmidt \cite{Schmidt}, Andersen and Sabio Vera \cite{AndersenSabioVera}, and Chachamis and Sabio Vera \cite{ChachamisSabioVera}, which separate emissions resolved above a small parameter from an exponentiated Regge form factor (Appendix~\ref{sec:supp-regularity}).

\section{The obstruction at next-to-leading order}
\label{sec:obstruction}

The next-to-leading kernel at fixed order contains a negative double logarithm of the ratio of transverse momenta. For strongly ordered momenta, this term removes emission probability faster than the leading term produces it. The candidate exponent at zero conformal spin in the symmetric scheme is the intercept minus the eigenvalue,
\begin{equation}
\Psi^{\rm NLO}(\nu)=\bar\alpha\big[\chi_0(\tfrac12)-\chi_0(\tfrac12+i\nu)\big]
+\bar\alpha^2\big[\chi_1(\tfrac12)-\chi_1(\tfrac12+i\nu)\big] ,
\label{eq:psinlo}
\end{equation}
where $\chi_1$ is the even part of the next-to-leading eigenvalue of Fadin and Lipatov \cite{FadinLipatov} and Ciafaloni and Camici \cite{CiafaloniCamici}, in the symmetric convention of Salam \cite{Salam}. Below, $\chi$ without a conformal-spin argument means $\bar\alpha\chi_0+\bar\alpha^2\chi_1$ at $n=0$. Thus $\Psi^{\rm NLO}(\nu)=\chi(\tfrac12)-\chi(\tfrac12+i\nu)$, while $\chi(n,\gamma)$ in \eqref{eq:chi} remains coupling-free.

The symmetric scheme fixes both the energy scale $s_0=q_1q_2$, where $q_1=k$ and $q_2=k'$, and the coupling at that argument. The scale alone leaves an odd term proportional to $\chi_0'$ from the running-coupling argument. The eigenfunction redefinition of Fadin and Lipatov \cite{FadinLipatov} cancels it. Evenness of $\chi$ under $\gamma\to1-\gamma$ requires both choices, as explained in Appendix~\ref{app:numerics}.

The reflection $\gamma\to1-\gamma$, under which $\chi$ is even in the symmetric scheme, also appears in identities for harmonic sums. Joubat and Prygarin \cite{JoubatPrygarin} use such identities to organize the most complicated terms of the next-to-next-to-leading eigenvalue in $\mathcal N=4$ super Yang-Mills at arbitrary conformal spin, and Joubat, Sandeu and Prygarin \cite{JoubatSandeuPrygarin} to organize powers of the leading-order eigenvalue at zero conformal spin.

A non-negative L\'evy density cannot generate \eqref{eq:psinlo} at any fixed positive coupling. An inequality near the edges of the strip $0<\mathrm{Re}\,\gamma<1$, where both eigenvalues are holomorphic, establishes the obstruction.

\subsection*{The strip-edge inequality}

A positive measure on $\mathbb G$ must have a positive radial projection, so zero conformal spin suffices for an exclusion. Suppose $\Psi$ is the exponent of a symmetric pure-jump process with L\'evy density $\pi\ge0$, and the corresponding $\chi$ continues to real arguments in $0<\mathrm{Re}\,\gamma<1$. We set $\gamma=\tfrac12+\sigma$, with $\sigma$ real. This shift of the argument weights each step $\delta$ by $e^{\sigma\delta}$ and the opposite step by $e^{-\sigma\delta}$, and the two weights average to $\cosh\sigma\delta\ge1$, so the eigenvalue can only rise away from the symmetric point,
\begin{equation}
\chi(\tfrac12)-\chi(\tfrac12+\sigma)=\int_{-\infty}^{\infty}
\big(1-e^{\sigma\delta}\big)\,\pi(|\delta|)\,d\delta
=\int_0^{\infty}\big(2-2\cosh\sigma\delta\big)\,\pi(\delta)\,d\delta\ \le\ 0 ,
\label{eq:stripedge}
\end{equation}
since $\pi\ge0$ fixes the sign in \eqref{eq:stripedge}. The first integral uses a symmetric cut at the origin, since it need not be absolutely convergent there. The paired half-lines give the expression on the right and the strip-edge inequality,
\begin{equation}
\chi(\tfrac12+\sigma)\ \ge\ \chi(\tfrac12)
\qquad\text{for every real }\sigma\text{ in the strip.}
\label{eq:stripineq}
\end{equation}
Equation \eqref{eq:stripineq} requires the symmetric point $\gamma=\tfrac12$ to be a minimum of the eigenvalue along the real direction. The next-to-leading combination tends to minus infinity near the edge of the strip, so it falls below $\chi(\tfrac12)$ there and violates \eqref{eq:stripineq}.

Convexity gives a stronger condition with fewer hypotheses. Let $\mu$ be a probability law on the line, with moment generating function $\Lambda(\sigma)=\int_{\mathbb R}e^{\sigma\delta}\,\mu(d\delta)$. A law generated by \eqref{eq:psinlo} at rapidity $Y$ would obey $\log\Lambda(\sigma)=Y\big[\chi(\tfrac12+\sigma)-\chi(\tfrac12)\big]$. The exponent $\Psi$ includes the coupling, so its time is $Y$, whereas \eqref{eq:transform} uses $t=\bar\alpha Y$. Two derivatives of \eqref{eq:LK} in $\sigma$ remove the killing, drift and truncation terms. The remaining terms are twice the Gaussian form $Q$ of \eqref{eq:LK} at unit radial frequency and the second moment of the tilted L\'evy measure, which is non-negative,
\begin{equation}
\chi''\big(\tfrac12+\sigma\big)=2Q(1)+\int_{\mathbb R\setminus\{0\}}
\delta^2e^{\sigma\delta}\,\Pi(d\delta)\ \ge\ 0 ,
\qquad
\chi(\tfrac12+\sigma)+\chi(\tfrac12-\sigma)-2\chi(\tfrac12)\ \ge\ 0 ,
\label{eq:convexity}
\end{equation}
so throughout the strip \eqref{eq:convexity} requires $\chi$ to be convex along the real direction.

The two inequalities hold even without a L\'evy-Khintchine representation. Finiteness of $\Lambda$ for $|\sigma|<\tfrac12$ is enough. The H\"older inequality makes $\log\Lambda$ convex, and Cauchy-Schwarz gives the midpoint inequality through
$1=\big(\int_{\mathbb R}e^{\sigma\delta/2}e^{-\sigma\delta/2}\,\mu(d\delta)\big)^2
\le\Lambda(\sigma)\Lambda(-\sigma)$.
Neither requires infinite divisibility, symmetry of a step measure, absolute continuity of $\Pi$, or the moment identity \eqref{eq:momentid}. A failure at one real $\gamma$ inside the strip excludes $\chi(\tfrac12)-\chi(\tfrac12+i\nu)$ as a characteristic exponent at every positive rapidity, since $Y$ is only a positive factor in $\log\Lambda$. The eigenvalue $\chi$ is even in $\sigma$ in the symmetric scheme, which reduces the midpoint condition to \eqref{eq:stripineq}, while convexity itself needs no evenness.

Ciafaloni, Colferai and Salam \cite{CCS} require positive curvature of their resummed $\omega$-dependent eigenvalue at its minimum near $\gamma=\tfrac12$, as a necessary condition for avoiding the oscillations found by Levin \cite{Levin} and by Ross \cite{Ross}. This requirement concerns the neighborhood of the minimum, whereas \eqref{eq:stripineq} and \eqref{eq:convexity} must hold at every point of the strip.

We write $\epsilon$ for the distance to the nearer edge of the strip, equal to $1-\gamma$ near $\gamma=1$ and to $\gamma$ near $\gamma=0$. The leading Laurent terms give $\chi''\simeq2\bar\alpha/\epsilon^3-6\bar\alpha^2/\epsilon^5$, which tends to $-\infty$ at the edge. Equation \eqref{eq:convexity} therefore fails at interior points at any fixed positive coupling. Holomorphy of $\chi_0$ and $\chi_1$ justifies the same continuation used in \eqref{eq:stripineq}. Near $\gamma\to1$, the eigenvalue itself behaves as
\begin{equation}
\bar\alpha\chi_0+\bar\alpha^2\chi_1
\;\simeq\;\frac{\bar\alpha}{\epsilon}-\frac{\bar\alpha^2}{2\epsilon^3} .
\label{eq:edgebehaviour}
\end{equation}
The cubic term has coefficient $d_{13}=\dthirteenval$ in $\chi_1$, as displayed in \eqref{eq:dvalues} below. The displayed terms cancel at $\epsilon=\sqrt{\bar\alpha/2}$, which is the leading-collinear estimate for the first failure of \eqref{eq:stripineq} at small coupling. Closer to the edge the cubic term dominates, so the combination diverges to $-\infty$, as stated after \eqref{eq:stripineq}. Appendix~\ref{sec:supp-obstruction} locates the first failures of the two inequalities at the smallest coupling of Figure~\ref{fig:nlo}, gives their limits at small coupling and compares those limits with the convergence radius of the square-root resummation of Salam \cite{Salam}.

The leading-collinear estimate of the first failure approaches the edge as the coupling decreases, but \eqref{eq:stripineq} still fails at any fixed positive coupling. Holomorphy of $\chi_0$ and $\chi_1$ makes $e^{-Y\Psi}$ holomorphic on the strip. The standard strip theorem for characteristic functions, as stated in the book of Lukacs \cite{Lukacs}, then implies that a probability law with this characteristic function would have finite $\Lambda$ throughout the strip. This finiteness is the hypothesis that the two inequalities \eqref{eq:convexity} need. Both fail, so $e^{-Y\Psi}$ is excluded as a characteristic function.

Each pole of the eigenvalue corresponds to a term in one exponential layer of the density. This correspondence follows from the half-line transform and its reflected partner, defined by
\begin{equation}
\hat T(\gamma)\equiv\int_0^\infty\big[e^{(\gamma-\frac12)x}-1\big]\,\pi(x)\,dx,
\qquad
\chi(\gamma)=\hat T(\gamma)+\hat T(1-\gamma)+{\rm const} ,
\label{eq:transformpair}
\end{equation}
where the two integrals converge together on $0<\mathrm{Re}\,\gamma<1$, and the relation continues to the rest of the plane. A pole of order $p$ at $\gamma=1+j$ corresponds to a polynomial of degree $p-1$ in $x$ in the $j$-th exponential layer,
\begin{equation}
\frac{K}{(1+j-\gamma)^p}\ \longleftrightarrow\
\pi(x)\supset\frac{K\,x^{p-1}}{(p-1)!}\,e^{-(j+\frac12)x} ,
\qquad j=0,1,2,\dots .
\label{eq:dictionary}
\end{equation}
The leading-order simple pole has unit residue and gives $\bar\alpha e^{-x/2}$, or $\bar\alpha$ times the $m=0$ layer of \eqref{eq:pin}. The next-to-leading Laurent expansion is
\begin{equation}
\chi_1(\gamma)=\frac{d_{13}}{(1-\gamma)^3}+\frac{d_{12}}{(1-\gamma)^2}
+\frac{d_{11}}{1-\gamma}+O(1),
\label{eq:dcoeffs}
\end{equation}
where the even eigenvalue in \eqref{eq:psinlo} has the coefficients given by Salam \cite{Salam},
\begin{equation}
d_{13}=-\frac12,\qquad
d_{12}=-\frac{11}{8}+\frac{n_f}{12N_c}\Big(1-\frac{2}{N_c^2}\Big),\qquad
d_{11}=-\frac{n_f}{36}\Big(\frac{10}{N_c}+\frac{13}{N_c^3}\Big) ,
\label{eq:dvalues}
\end{equation}
so the cubic pole contributes $-\bar\alpha^2x^2e^{-x/2}/4$. Only $d_{13}$ is independent of $N_c$ and $n_f$, so the obstruction holds for every $N_c$ and $n_f$, although the displacement at which the density crosses zero depends on them.

The dictionary \eqref{eq:dictionary} relates the double logarithm described at the start of this section to the cubic pole. A pole of order $p$ at $\gamma=1$ gives a logarithm of order $p-1$ in the momentum ratio. The simple pole gives the leading-order collinear enhancement, and the cubic pole produces the negative double logarithm $\ln^2(k'^2/k^2)$. Salam \cite{Salam} and Ciafaloni and Colferai \cite{CiafaloniColferai} resum this logarithm. At fixed order, the same coefficient $d_{13}$ in \eqref{eq:dvalues}, independent of $N_c$ and $n_f$, causes both the intercept instability and the loss of positivity of the step measure.

\subsection*{The local form, and the coupling it produces}

The second-order expansion of \eqref{eq:stripedge} in $\sigma$ tests the second moment of the step measure. The finiteness of $\Lambda$ on the strip, which the strip theorem gives \cite{Lukacs}, justifies differentiation under the integral through the equivalence, given by Sato \cite{Sato} for an infinitely divisible law, between this finiteness and finiteness of the same exponential integral against the L\'evy measure away from the origin (Appendix~\ref{sec:supp-obstruction}). The even derivatives of the eigenvalue at the symmetric point are the even moments of a pure-jump density $\pi\ge0$,
\begin{equation}
\chi^{(2j)}(\tfrac12)=\int_{-\infty}^{\infty}\delta^{2j}\,\pi(|\delta|)\,d\delta ,
\qquad j=1,2,\dots ,
\label{eq:momentid}
\end{equation}
and every such moment must be non-negative. The expansion of $\cosh\sigma\delta$ in \eqref{eq:stripedge} in powers of $\sigma$ gives the identity term by term, since the exponent is even in $\sigma$ and the moments are finite. Equivalently, $2j$ derivatives of the exponent bring down $\delta^{2j}$, and the alternating signs cancel at $\nu=0$.

No inverse transform or continuation beyond the strip is needed. At \mbox{$j=1$}, the identity reads \eqref{eq:secondmoment} in reverse. For \eqref{eq:psinlo} at $j=1$, a positive step measure requires the diffusion coefficient \mbox{$\chi''(\tfrac12)=\bar\alpha\chi_0''(\tfrac12)+\bar\alpha^2\chi_1''(\tfrac12)$} to be non-negative. At $N_c=3$ and $n_f=4$, this requirement fails once
\begin{equation}
\bar\alpha\ >\ \bar\alpha_c=\frac{28\zeta_3}{|\chi_1''(\tfrac12)|}=\abarcrit ,
\qquad \chi_0''(\tfrac12)=28\zeta_3=\chizeropp,\qquad
\chi_1''(\tfrac12)=\chionepp .
\label{eq:abarc}
\end{equation}
The coupling in \eqref{eq:abarc} marks the zero of the diffusion coefficient, where the second moment changes sign. Above $\bar\alpha_c$, the second-moment condition fails at the symmetric point itself, so \eqref{eq:stripineq} fails in every neighborhood of that point. The strip-edge argument already excludes positivity below $\bar\alpha_c$.

Bl\"umlein and Vogt \cite{BlumleinVogt} found a related sign reversal in the then available irreducible contribution to the resummed gluon anomalous dimension. Its subleading small-$x$ coefficients are mostly opposite in sign to the leading ones and are typically larger in magnitude in Mellin space. Levin \cite{Levin} uses the expansion coefficients of Ross \cite{Ross} to locate the diffusion sign change. The text below Eq.~(2.13) of Levin \cite{Levin} gives $\bar\alpha_S\approx0.05$, compared with $\abarcrit$ here. Equation \eqref{eq:momentid} identifies this change with a negative second moment of the would-be step measure.

The layer representation shows how the cubic pole causes this failure. Set $x=|\delta|$, the absolute radial displacement. Equation \eqref{eq:dictionary} maps a pole of order $p$ at $\gamma=1$ to a polynomial of degree $p-1$ times $e^{-x/2}$. The cubic contribution $-\bar\alpha^2x^2e^{-x/2}/4$ eventually outweighs the positive leading-order layer $\bar\alpha e^{-x/2}$. Figure~\ref{fig:nlo} shows the crossing obtained by inversion. The estimated positivity interval in Appendix~\ref{app:remainder} ends before it. The strip-edge and moment proofs are independent of this inversion and of the figure.

\begin{figure}[!b]
\centering
\includegraphics[width=0.85\textwidth]{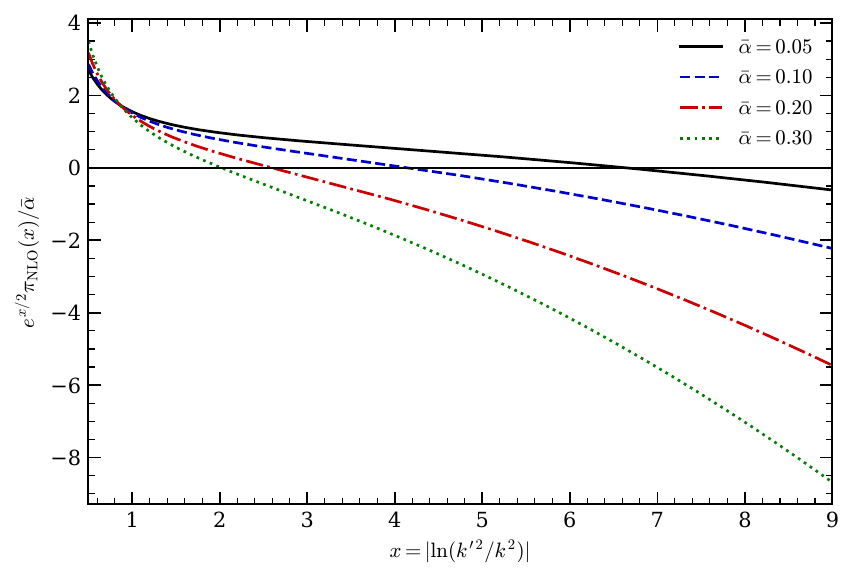}
\caption{The inverted fixed-order next-to-leading density of \eqref{eq:psinlo}
at $N_c=3$ and $n_f=4$. The ordinate is $e^{x/2}\pi_{\rm NLO}(x)/\bar\alpha$
and the abscissa is $x=|\ln(k'^2/k^2)|$, with the four couplings given in the
legend. The rescaling by $e^{x/2}$ and by $\bar\alpha$ makes the leading-order layer common
to the four curves and places them on the same axis. Each curve is
positive at small $x$ and becomes negative when the term of order
$\bar\alpha^2$ equals minus the term of order $\bar\alpha$. The crossing
locates the failure of the truncated density, not a physical
transverse-momentum scale.}
\label{fig:nlo}
\end{figure}

\subsection*{The limits of the obstruction}

First-order truncation can destroy positivity even when the full density is positive. The control density $\pi_\alpha(x)=\pi_0(x)\,e^{-\alpha x^2/4}$ is a L\'evy density for every $\alpha>0$, and its exponent is holomorphic in $\mathrm{Re}\,\alpha>0$. The first-order truncation of the exponent in $\alpha$ gives $\pi_0(x)\big(1-\alpha x^2/4\big)$, which becomes negative beyond $x=2/\sqrt{\alpha}$, where the large-$x$ correction overtakes the positive leading term. The even-moment thresholds of this truncated density tend to zero. The control density contains no QCD. Appendix~\ref{sec:supp-obstruction} derives these thresholds and contrasts the control density with a bounded correction, whose truncation stays positive up to a finite value of its parameter.

The higher even-moment conditions for the fixed-order BFKL exponent likewise give coupling thresholds that decrease to zero. This decrease to zero is a second route to the exclusion at any fixed positive coupling. These thresholds reflect first-order truncation, as do those of the control density. Their law and first values are given in Appendix~\ref{sec:supp-obstruction}.

The Laurent data also permit a positive radial completion at zero conformal spin. Let $(K_{3j},K_{2j},K_{1j})$ be the coefficients of $\chi_1$ at $\gamma=1+j$ in \eqref{eq:dictionary}. Here $j$ counts layers, and $K_{3j}=d_{13}$ in every layer. The completion exponentiates each layer polynomial,
\begin{equation}
\pi_{\rm c}(x)=\bar\alpha\sum_{j\ge0}e^{-(j+\frac12)x}\,
\exp\Big[\bar\alpha\Big(\tfrac12K_{3j}x^2+K_{2j}x+K_{1j}\Big)\Big] .
\label{eq:completion}
\end{equation}
Every factor in \eqref{eq:completion} is positive, so $\pi_{\rm c}\ge0$. The expansion of the exponential reproduces the truncated density through order $\bar\alpha^2$.

We write $\bar b=(11N_c-2n_f)/(12N_c)$ for the first coefficient of the beta function divided by $4N_c$. Its role in the eigenvalue is given in Appendix~\ref{app:numerics}. The completion is a L\'evy density for $\bar\alpha<1/\bar b$ and has finite variation for $\bar\alpha<1/(2\bar b)$ (Appendix~\ref{sec:supp-obstruction}). Both bounds lie well above the largest coupling used, $\bar\alpha=\scanabarD$, since $\bar b=\bbarvalue$ at $N_c=3$ and $n_f=4$. At $N_c=3$, the bound for a L\'evy density reads $\alpha_s<12\pi/(33-2n_f)$, which exceeds one for every $n_f\le6$, so the completion is a L\'evy density for every QCD coupling below one.

This completion concerns the normalized radial exponent at zero conformal spin. It does not specify an all-spin kernel or derive the kernel beyond the computed order. It shows that the perturbative data allow a positive radial L\'evy density, although \eqref{eq:psinlo} itself has none. The obstruction is therefore a property of the fixed-order truncation. The control density and the next-to-leading kernel can be distinguished only through their resummations. Exponentiation of the quadratic term restores the positive control density $\pi_\alpha$ for every $\alpha$. Section~\ref{sec:removal} tests whether the collinear resummations in use restore positivity for the BFKL kernel.

\section{Conjugations and collinear resummations}
\label{sec:removal}

The negative fixed-order density could in principle be avoided by changing the conjugation \eqref{eq:conjugation} or by resumming the perturbative series. We first show that continuous pointwise multiplicative conjugations preserving translations cannot change its sign. We then test the symmetric-scheme resummations, the double-logarithmic kernel of Iancu et al.\ \cite{IMMST}, and the anti-collinear kernel of Kovner, Lublinsky, Nefedov and Skokov \cite{KLNSone} in a different rapidity-factorization scheme.

The tests do not cover running-coupling resummations. Examples are the resummation of Altarelli, Ball and Forte \cite{ABF}, implemented in HELL by Bonvini, Marzani and Peraro \cite{HELL}, and that of Thorne \cite{Thorne}. Altarelli, Ball and Forte resum logarithms of $1/x$, collinear logarithms and running-coupling logarithms. Thorne resums those of the running coupling and $1/x$, leaving collinear resummation aside. A running coupling removes translation invariance in $\ln k^2$, which the derivations of \eqref{eq:stripineq} and \eqref{eq:convexity} use.

\subsection*{The family of admissible conjugations}

The choice $g=k f$ makes the step measure symmetric. Translation invariance alone allows $g=k^{s}f$ for every $s$, because the kernel remains homogeneous of degree zero. Finiteness of the tilted measure away from the origin also requires $|1-s|<1$. Thus $s$ lies in $0<s<2$, keeping the Mellin contour in $0<\mathrm{Re}\,\gamma<1$. Only $s=1$ gives a symmetric measure. It selects the line $\mathrm{Re}\,\gamma=\tfrac12$, and this choice amounts to choosing a frame.

Changing the conjugation shifts the line on which the exponent is evaluated,
\begin{equation}
\Psi_s(\nu,n)=\chi\big(0,\tfrac12+\theta_{\rm E}\big)
-\chi\big(n,\tfrac12+\theta_{\rm E}-i\nu\big),
\qquad \theta_{\rm E}=\frac{1-s}{2} ,
\label{eq:tilted}
\end{equation}
which is an Esscher transform. It multiplies the density by $e^{\theta_{\rm E}\delta}$, adjusts the normalization, and moves the contour to $\mathrm{Re}\,\gamma=\tfrac12+\theta_{\rm E}$. Inverting on that line gives
\begin{equation}
\rho_s(\delta,\phi)=e^{\theta_{\rm E}\delta}\,\rho(\delta,\phi) .
\label{eq:esscher}
\end{equation}
The exponential factor is positive and cannot change the sign at any step. Holomorphy of $\chi_1$ inside the strip also prevents the contour shift from moving the cubic pole. The family $g=k^{s}f$ exhausts the continuous pointwise multiplicative conjugations that keep the kernel translation invariant and the density real and non-negative (Appendix~\ref{sec:supp-resummation}). Every admissible conjugation therefore retains the fixed-order obstruction.

A change of rapidity-factorization scheme is a different operation from a conjugation. It moves $s_0$ while keeping the coupling argument fixed as in Section~\ref{sec:obstruction}. Either asymmetric choice relative to $s_0=q_1q_2$ cancels the cubic pole at one edge and doubles it at the other. The resulting $\chi$ is no longer even in $\sigma$, so \eqref{eq:stripineq} does not apply. Convexity \eqref{eq:convexity} remains applicable and excludes both asymmetric fixed-order choices, as shown below.

\subsection*{The collinear shift}

We work at zero conformal spin in this and the next subsection. The radial projection of Section~\ref{sec:obstruction} then makes a test of this one Fourier coefficient sufficient to exclude a non-negative cylinder density. The resummation of Salam \cite{Salam} and Ciafaloni and Colferai \cite{CiafaloniColferai} shifts each leading-order collinear pole by $\omega/2$. Inverting each shifted pole gives
\begin{equation}
\pi_\omega(x)=\bar\alpha\,\pi_0(x)\,e^{-\omega x/2}\,\theta(x) ,
\label{eq:shiftedtower}
\end{equation}
which is non-negative for $\omega\ge0$ and $x>0$, with $\theta(x)$ the Heaviside function. The shift acts as a delay in the rapidity variable conjugate to $\omega$, and a delay preserves the transverse sign, so the obstruction does not arise from the shift itself. The sign change arises when $\omega$ is eliminated to obtain a rapidity-local exponent.

\subsection*{Resumming every collinear pole}

Sabio Vera \cite{SabioVera} solves the shift at each pole and sums the solutions in the all-poles approximation. The single-pole solution is $\bar\alpha$ times the function $\chi_L$ of Salam \cite{Salam}. Including the virtual subtraction inside the sum gives the convergent eigenvalue
\begin{equation}
\omega(\gamma)=\sum_{m\ge0}\Big[F(\gamma+m)+F(1-\gamma+m)
-\frac{2\bar\alpha}{m+1}\Big],\qquad
F(\lambda)=\sqrt{\lambda^2+2\bar\alpha}-\lambda ,
\label{eq:allpoles}
\end{equation}
where the expansion $F(\lambda)=\bar\alpha/\lambda-\bar\alpha^2/(2\lambda^3)
+O(\bar\alpha^3)$ has cubic coefficient $-\tfrac12$ in every layer. The family in \eqref{eq:allpoles} near $\gamma=1$ is $F(\epsilon+m)$, whose $m$-th member resums the pole at $\gamma=1+m$. Inverting all layers gives
\begin{equation}
\pi_{\rm eff}(x)=\pi_0(x)\,\sqrt{2\bar\alpha}\,
\frac{J_1(\sqrt{2\bar\alpha}\,x)}{x} ,
\label{eq:bessellayer}
\end{equation}
where the multiplier of $\pi_0$ is the Bessel factor of the all-poles density. Its single-layer transform is
\begin{equation}
\int_0^\infty e^{-\lambda x}\,\frac{a\,J_1(ax)}{x}\,dx=\sqrt{\lambda^2+a^2}-\lambda ,
\qquad a=\sqrt{2\bar\alpha} ,
\label{eq:besselpair}
\end{equation}
so the function inverted from $F$ begins at $\bar\alpha$, as required by the leading-order layer, but oscillates at larger distances. Summing the layers with \eqref{eq:pin} gives $1/(1-e^{-x})$ and yields \eqref{eq:bessellayer}.

The density \eqref{eq:bessellayer} is the integrand resummed by Sabio Vera \cite{SabioVera}. The Bessel factor appears both in the integral representation of the all-poles eigenvalue and in the associated real-emission kernel. The cure of the Green-function oscillations and positivity of the step density are different requirements. The fixed-order Green function oscillates and becomes negative when $q_1^2/q_2^2$ is far from unity. Chachamis and Sabio Vera \cite{ChachamisSabioVera} also find a negative Green function in the collinear region when the leading-order iterative solution keeps only the leading collinear double logarithm. The Bessel oscillation in the resummation compensates for the oscillations of the fixed-order Green function, as described by Sabio Vera \cite{SabioVera}.

The step density \eqref{eq:bessellayer} retains the Bessel zeros because $\pi_0>0$. Its first sign change is at $\sqrt{2\bar\alpha}\,x=j_{1,1}$, or $x^{\rm
B}_*=\bessconst/\sqrt{\bar\alpha}$, and it changes sign infinitely often thereafter. The origin of this oscillation is the square-root branch point. The function $F(\lambda)=\sqrt{\lambda^2+2\bar\alpha}-\lambda$ has a branch point at $\lambda=i\sqrt{2\bar\alpha}$. Salam \cite{Salam} describes its approach to $\gamma=0$ and $\gamma=1$ as the coupling decreases. The inverse-transform scale is $x^{\rm B}_*$, so the resummed square root itself generates the sign change, without a fixed-order truncation.

Next-to-leading matching sets the two constants of the all-poles approximation of Sabio Vera \cite{SabioVera}, both zero in the pure form above, from $d_{11}$ and $d_{12}$ of \eqref{eq:dvalues}. Matching multiplies the pure density by a positive factor and changes its Bessel argument through $d_{11}$. The matched density is negative somewhere at every positive coupling, except at $\bar\alpha=1/|d_{11}|$ for $n_f\ne0$, a coupling, far outside the physical range, at which the step measure vanishes and the process never jumps (Appendix~\ref{sec:supp-resummation}).

At $N_c=3$ and $n_f=4$, the full prescription of Sabio Vera \cite{SabioVera} adds to the matched density a correction that is positive for every $x>0$, but the full density still changes sign at both couplings computed, and the conclusion for this form is restricted to them (Appendix~\ref{sec:supp-resummation}). The failure of the pure and matched forms, by contrast, follows in closed form, apart from the coupling at which the matched step measure vanishes.

A single-layer test also excludes a non-negative measure. The Bernstein theorem \cite{BergForst} makes a non-negative layer density equivalent to complete monotonicity of its transform $F$, that is, to derivatives of alternating sign. Its fourth derivative
\begin{equation}
F''''(\lambda)=-\frac{12\,\bar\alpha\,(\bar\alpha-2\lambda^2)}{(\lambda^2+2\bar\alpha)^{7/2}}
\label{eq:cmfail}
\end{equation}
is negative for $\lambda<\sqrt{\bar\alpha/2}$, violating complete monotonicity at every positive coupling within $\sqrt{\bar\alpha/2}$ of the pole. This is the scale in \eqref{eq:edgebehaviour} where the term of order $\bar\alpha^2$ overtakes that of order $\bar\alpha$. The rapidity-local resummation \eqref{eq:allpoles} therefore has no non-negative step measure. The positive collinear result of Chachamis and Sabio Vera \cite{ChachamisSabioVera} concerns the finite-rapidity Green function after restoring the collinear terms, not the intercept-subtracted generator tested here.

The square root also occurs outside the symmetric scheme. Iancu et al.\ \cite{IMMST} resum double transverse logarithms of the next-to-leading Balitsky-Kovchegov kernel to all orders in the double-logarithmic approximation. Their rapidity-local exponent is $\bar\alpha\chi_{\rm
DLA}(\gamma)=\tfrac12\big[\sqrt{(1-\gamma)^2+4\bar\alpha}-(1-\gamma)\big]$. It equals $\tfrac12 F(1-\gamma)$ of \eqref{eq:allpoles} at twice the coupling, keeping only the $m=0$ collinear layer. Their kernel is the integrand of \eqref{eq:besselpair} with $a$ replaced by $2\sqrt{\bar\alpha}$, divided by $2\bar\alpha$, and it changes sign at $2\sqrt{\bar\alpha}\,x=j_{1,1}$. The same replacement in \eqref{eq:cmfail} gives a negative derivative for $\lambda<\sqrt{\bar\alpha}$, so complete monotonicity fails within $\sqrt{\bar\alpha}$ of the pole. The positive solution they report applies in the region where the local formulation reproduces the non-local equation, and not to the step measure, which changes sign (Appendix~\ref{sec:supp-resummation}).

\subsection*{The resummation that removes the pole}

Another resummation removes the pole of $\chi_0$ at $\gamma=1$ altogether. Kovner, Lublinsky, Skokov and Zhao \cite{KLSZ} identify the terms producing DGLAP-like transverse logarithms in the next-to-leading JIMWLK kernel. They derive an evolution resumming them to all orders in $\alpha_s$, with approximate solutions in the dilute and saturated regimes. Kovner, Lublinsky, Nefedov and Skokov \cite{KLNSone} study the linear fixed-coupling regime and give the resummed kernel and eigenvalue in closed form. At zero spin, the eigenvalue remains finite as $\gamma\to1$.

They call $\gamma=1$ the anti-collinear edge, whereas we call it collinear. The choice depends on which scatterer is the projectile. The correction acts only at zero spin and has all poles at $\gamma>0$, so the reflected term of \eqref{eq:transformpair} does not contribute. Inversion at $n_f=0$ gives the radial correction
\begin{equation}
\Delta\pi(x)=-e^{-x/2}\big(1-e^{-\lambda_{\rm sc}x}\big),\qquad
\lambda_{\rm sc}=\frac{\bar\alpha\beta_0}{4N_c} ,
\label{eq:acdensity}
\end{equation}
where $\beta_0=(11N_c-2n_f)/3$ is the first beta-function coefficient, giving $\lambda_{\rm sc}=11\bar\alpha/12$ at zero flavors. At nonzero flavor number, two gluon-quark mixing rates replace this rate and sum to $\alpha_s(11N_c+2n_f)/(12\pi)$.

The correction \eqref{eq:acdensity} is uniform in azimuth, so the cylinder density is smallest at $\phi=\pi$. There \eqref{eq:rho} equals $r/[2\pi\,(1+r)^2]$, in terms of the wrapped Cauchy parameter $r$ of \eqref{eq:wrappedcauchy}. Positivity reduces to $r^{2\lambda_{\rm
sc}-1}(1+r)^2\ge2+r$ on $0<r<1$. It holds at all four tested couplings for $n_f=0$, with the least margin at the largest coupling, and first fails at $\bar\alpha=\acabarzero$. The corresponding two-exponential condition at three and four flavors also holds at all four tested couplings (Appendix~\ref{sec:supp-resummation}).

This leading-logarithmic resummation uses the $(+)$ rapidity-factorization scheme. The obstruction, by contrast, concerns the next-to-leading eigenvalue in the symmetric scheme. The relation between the two schemes is a scale change, distinct from \eqref{eq:conjugation}. The change shifts $\chi_1$ by $\pm\tfrac12\chi_0\chi_0'$, with the upper sign for $(+)$. The term $\chi_0\chi_0'$ approaches $\epsilon^{-3}$ at the collinear edge, so the shift cancels the cubic pole there, as stated in Ref.~\cite{KLNSone} and reflected in the resummed eigenvalue given there.

The fixed-order exponent $\bar\alpha\chi_0+\bar\alpha^2\chi_1^{(+)}$, with $\chi_1^{(+)}=\chi_1+\tfrac12\chi_0\chi_0'$, still fails \eqref{eq:convexity} at $N_c=3$ and $n_f=4$. The shift removes $d_{13}$ at $\gamma\to1$ but leaves $d_{12}$ of \eqref{eq:dvalues} unchanged. More generally, a scale change by any real power of $q_1/q_2$ shifts the two edge cubic coefficients equally and oppositely, leaving their sum $2d_{13}<0$, so convexity fails at a collinear edge for every such scale and every positive coupling at fixed order. Appendix~\ref{sec:supp-resummation} locates the convexity failures in both asymmetric schemes and gives this argument for every power.

\subsection*{The improved Green function at finite rapidity}

An $\omega$-dependent kernel requires inversion at finite rapidity. It cannot be inserted as an exponent. The resolvent is
\begin{equation}
\mathcal R(\omega,\gamma)=\frac{1}{\omega-\bar\alpha X(\omega,\gamma)},
\qquad X=X_0+\bar\alpha X_1 ,
\label{eq:rgiGreen}
\end{equation}
where the term $X_0$ contains the leading-order shift of Ciafaloni, Colferai and Salam \cite{CCS} and Ciafaloni, Colferai, Salam and Sta\'sto \cite{CCSS}. The coefficient $X_1$ is the improved next-to-leading coefficient of Colferai, Li and Sta\'sto \cite{ColferaiLiStasto}, used in their resummed virtual photon-photon cross section. Their five prescriptions, collA, collB, zVnB, zVnM and zVzM, differ through $U(\omega)$ in $X_0$. The last three set the double-pole coefficient $V$ in $X_1$ to zero and share the same $U$, giving one Green function. We therefore test three distinct prescriptions from this one family.

Let $P_t(\nu)$ be the zero-spin inverse transform of \eqref{eq:rgiGreen} at fixed rapidity, normalized to $P_t(0)=1$,
\[
P_t(\nu)\ \propto\ \frac{1}{2\pi i}\int_{\mathrm{Re}\,\omega=\omega_0}
e^{\omega Y}\,\mathcal R\big(\omega,\tfrac12+i\nu\big)\,d\omega ,
\]
where the abscissa $\omega_0$ must lie to the right of every singularity of $\mathcal R$ in $\omega$. Normalization fixes the $\nu$-independent factor, which is positive at every prescription, coupling and time tested.

Positive definiteness of $P_t$ as a function of frequency is weaker than the existence of a L\'evy density. The Bochner theorem \cite{BergForst} identifies positive definiteness of continuous $P_t$ with being the transform of a positive measure. In particular, every finite Toeplitz matrix with entries $P_t(\nu_j-\nu_k)$ at equally spaced grid points $\nu_j$ must have non-negative eigenvalues. Giraud and Peschanski \cite{GiraudPeschanski} use this test for phenomenological dipole amplitudes, and Peschanski \cite{PeschanskiTMD} for transverse-momentum-dependent gluon distributions, in both cases for what scatters and not for an evolution kernel.

The inverse transform $P_t$ is computed with the vertical-contour method of de Hoog, Knight and Stokes \cite{deHoog}. The contour assumption is that every singularity of \eqref{eq:rgiGreen} lies to the left of this contour. An argument-principle count of roots in Appendix~\ref{sec:supp-finite} supports this assumption, but the count is a computation, not a proof, and the finite-rapidity conclusions depend on it.

The five-point Toeplitz matrix has a negative smallest eigenvalue for all three prescriptions of this one family at all four tested couplings, from $\bar\alpha=\scanabarA$ to $\scanabarD$, for $N_c=3$ and $n_f=4$ at the shortest of the five times, $t=\scantA$, subject to the contour assumption. At later times fewer combinations give a negative eigenvalue. The eigenvalue tests apply separately to each tested combination of prescription, coupling and time, on two frequency grids, and a negative eigenvalue excludes positive definiteness of $P_t$ for that combination, whereas a non-negative one leaves it unresolved. The kernel terms are included one at a time, and the negative eigenvalue appears with the $\omega$-shift and $U(\omega)$ on the five-point grid at $\bar\alpha=\scanabarA$ and $t=\scantA$, before $X_1$ enters. The sign failure there is therefore a property of the resummed kernel itself. The frequency grids, the couplings, times and rapidities, the smallest eigenvalues, the stage-by-stage eigenvalues, the bound $|P_t|\le1$ and the semigroup test are given in Appendix~\ref{sec:supp-finite}, and a coefficient vector realizing the smallest violation, with its error estimate, in Appendix~\ref{app:numerics}.

The hypotheses distinguish the different exclusions of this paper. The strip-edge inequality \eqref{eq:stripineq} requires a rapidity-local generator, translation invariance in $\ln k^2$, holomorphy on $0<\mathrm{Re}\,\gamma<1$, and evenness under $\gamma\to1-\gamma$. The local double-logarithmic exponent of Iancu et al.\ \cite{IMMST} depends only on $1-\gamma$ and is not even in $\sigma$. Its second derivative $2\bar\alpha\big[(1-\gamma)^2+4\bar\alpha\big]^{-3/2}$ is positive on the strip, so this exponent satisfies the convexity condition \eqref{eq:convexity}, which needs no evenness. Convexity \eqref{eq:convexity} is necessary but insufficient. The single-layer complete-monotonicity test above excludes a non-negative step measure for this exponent, using the square root of \eqref{eq:allpoles} at twice the coupling. The exclusions on the cylinder also hold if the same kernel is represented on a larger state space with the same cylinder projection or by rapidity-dependent translation-invariant L\'evy generators (Appendix~\ref{sec:supp-resummation}).

\section{Discussion}
\label{sec:discussion}

We have shown that the forward BFKL equation at leading order and fixed coupling defines a L\'evy process with no approximation, once the gluon density is rescaled by the magnitude of its transverse momentum and its growth is divided out. A single positive step measure on the cylinder of logarithmic momentum and azimuth covers every conformal spin and rapidity, and the same measure gives the intercept, the diffusion and the azimuthal relaxation. The process is symmetric under step inversion and consists only of jumps, infinitely many of them short, with a finite total length over any finite range of rapidity. We have also shown that the angle-averaged evolution truncated at next-to-leading order in the symmetric scheme admits no probability law at any fixed positive coupling and rapidity. Near the collinear poles, the eigenvalue falls below its value at the symmetric point, through a negative double logarithm whose coefficient is independent of the numbers of colors and flavors. A positive radial completion at zero conformal spin agrees with the truncated density to the computed order and is a L\'evy density for every QCD coupling $\alpha_s<1$ (Section~\ref{sec:obstruction}), so the failure lies in the truncation. The main message of this paper is that a probability law for the random walk of forward evolution at fixed coupling exists at leading order but not for the fixed-order next-to-leading truncation in the symmetric scheme, so matching the perturbative expansion is not enough. A resummed kernel meant to describe a random walk must also have a positive step measure, which at fixed coupling can be tested on the kernel itself.

In Section~\ref{sec:removal} we test whether the obstruction can be removed. No reweighting of the density by a power of the transverse momentum removes it, and these are the only continuous reweightings that keep the kernel translation invariant and the density real and non-negative. The all-poles approximation of Sabio Vera \cite{SabioVera} resums the collinear shift at every pole of the leading-order eigenvalue. We show in closed form that its step density takes negative values at every positive coupling in the pure form and with next-to-leading matching, apart from one coupling at which the matched step measure vanishes. The step density of the full prescription also takes negative values at the two couplings computed at four flavors. The collinearly improved Green functions of Colferai, Li and Sta\'sto \cite{ColferaiLiStasto} are not transforms of probability distributions at any of the four couplings tested at the shortest of the five evolution times, if every singularity of these Green functions lies to the left of the inversion contour (Section~\ref{sec:removal}). The leading-logarithmic resummed kernel of Kovner, Lublinsky, Nefedov and Skokov \cite{KLNSone}, which removes the collinear pole of the leading-order eigenvalue in another rapidity-factorization scheme, gives a non-negative step density at the couplings and flavor numbers tested. Kernels with a running coupling lie outside the derivations, which use translation invariance in $\ln k^2$ (Section~\ref{sec:removal}).

The leading-order process permits unweighted Monte Carlo sampling of all steps larger than any radial cut, whose rate is finite by Eq.~\eqref{eq:activity}, while the mean distance accumulated by the shorter steps vanishes as the cut is removed, since the mean accumulated distances \eqref{eq:moments} are finite. Schmidt \cite{Schmidt} gives a leading-order iterative series with positive terms, whereas the next-to-leading emission factor studied by Chachamis and Sabio Vera \cite{ChachamisSabioVera} changes sign and requires signed contributions. In zero transverse dimensions, negative Abramovskii-Gribov-Kancheli weights also produce a sign problem, which Ouchen and Prygarin \cite{OuchenPrygarin} remove by a Euclidean reformulation. The fixed-order next-to-leading kernel in the symmetric scheme has no positive step measure at any fixed positive coupling, so a passage to nonlinear evolution justified by its stochastic interpretation loses that justification, without invalidating the nonlinear evolution equations. Its normalized Green function is likewise not a probability distribution at any fixed positive coupling and rapidity, so it has no Shannon entropy as one, whereas the entropy of Kharzeev and Levin \cite{KharzeevLevin} is unaffected. That entropy equals the von Neumann entropy of a reduced density operator diagonal in the multiplicity basis and concerns a different distribution, the multiplicity distribution of the cascade, which has a leading-order probability law.

Related positivity failures arise in the next-to-leading kernel of the JIMWLK equation, which describes how scattering off a dense system of gluons evolves with energy. Altinoluk, Beuf, Lublinsky and Skokov \cite{ABLS} require the coefficient of its leading-order color operator, as a quadratic form in two transverse separations, to be positive semidefinite for a Langevin description. They find that the fixed-order kernel and conventional running-coupling prescriptions violate this condition, which is of the same type as the positive-definiteness test of Section~\ref{sec:removal}, and leave open how to test the whole next-to-leading Hamiltonian. The daughter-dipole prescription factorizes and satisfies the condition, and their analysis keeps terms proportional to the first beta-function coefficient. Kovner, Lublinsky, Skokov and Zhao \cite{KLSZ} place terms outside that prescription into DGLAP evolution of the projectile, and Section~\ref{sec:removal} tests the linear fixed-coupling form of their resummation, given by Kovner, Lublinsky, Nefedov and Skokov \cite{KLNSone}.

Lappi and M\"antysaari \cite{LappiMantysaari} find negative dipole amplitudes at small sizes for some initial conditions in next-to-leading Balitsky-Kovchegov evolution. They trace the instability to the double transverse logarithm. Iancu et al.\ \cite{IMMST} stabilize the evolution by resumming the double transverse logarithm while leaving finite next-to-leading terms aside, and whether this instability and the forward-kernel obstruction are one failure seen in different variables is not known. The open problem for the symmetric next-to-leading kernel is whether a collinear resummation gives a non-negative step measure outside the forms and parameter values that Section~\ref{sec:removal} excludes. Those forms are based on the resummations of Salam \cite{Salam}, of Ciafaloni, Colferai and Salam \cite{CCS}, of Sabio Vera \cite{SabioVera} and of Colferai, Li and Sta\'sto \cite{ColferaiLiStasto}. The radial completion shows that the perturbative data allow a non-negative step measure at zero conformal spin, but we neither construct a resummed kernel with this property nor exclude its existence.

We propose testing any resummed fixed-coupling kernel intended for a Monte Carlo program for positivity before it is used to generate events, by checking that its eigenvalue along the real axis between the collinear poles is convex and, in a symmetric scheme, smallest at the symmetric point, and that the transform of each layer is completely monotonic. These conditions apply if the kernel is rapidity-local and translation invariant in $\ln k^2$, with an exponent holomorphic on the strip of Section~\ref{sec:obstruction}. A first target is the full prescription in the form that Chachamis and Sabio Vera \cite{ChachamisSabioVera} implement in a Monte Carlo event generator, at zero flavors and without the terms proportional to the first beta-function coefficient, whereas Section~\ref{sec:removal} tests this prescription with these terms at four flavors and two couplings.

The main result of this paper is that the gluon density of forward BFKL evolution, multiplied by the magnitude of its transverse momentum and divided by its growth, evolves at leading order and fixed coupling as a L\'evy process with no approximation, and that one positive measure gives the intercept, the diffusion and the azimuthal relaxation of this evolution. We have also shown that the fixed-order next-to-leading truncation of the angle-averaged evolution in the symmetric scheme has no probability law at any fixed positive coupling and rapidity, a failure of the truncation, since a positive radial completion at zero conformal spin agrees with the truncated density to the computed order and is a L\'evy density for every QCD coupling $\alpha_s<1$. The main message is that agreement with the perturbative expansion does not by itself give positive step probabilities to the random walk of forward evolution at fixed coupling. Positivity is therefore a separate requirement for a resummed kernel meant to describe a random walk, and it can be tested at fixed coupling before the kernel is used.

\section*{Data availability statement}
The original contributions are included in this article. The programs and stored results that reproduce every number and both figures are publicly available as ancillary files of the arXiv preprint of this work, arXiv:2609.18269. Further inquiries should be directed to the corresponding author.

\section*{Acknowledgements}
We thank Sergey Bondarenko for fruitful discussions.

\subsection*{AI disclosure statement}
The authors formulated the physics problem and directed the work at every stage. Generative AI models carried out the technical execution under their direction, including the derivations and calculations, the programs producing the numbers and both figures, independent recomputations and cross-checks, and the drafting of the text. The models were Claude Opus 5, Claude Opus 5.5 and Claude Fable 5.1 from Anthropic, GPT-5.6 Sol and GPT-6 Astra from OpenAI, Gemini 3.1 Pro from Google, and Kimi K3 from Moonshot AI.

The authors specified the computational tasks and checking criteria, ran and inspected the programs, and checked the reported results. The co-authors checked the derivations and numbers. Every number and both figures are reproduced by the programs in the public archive identified in the Data availability statement. The figures were drawn by programs written by these models, and the authors checked them against the computed data. The authors take full responsibility for the content.

\appendix
\numberwithin{equation}{section}

\section{Derivation of the characteristic exponent}
\label{app:exponent}

We obtain the exponent by transforming each layer of the density. Angular integration uses the Poisson-kernel identity
\begin{equation}
\int_{-\pi}^{\pi}\frac{e^{-in\phi}\,d\phi}{a-\cos\phi}
=\frac{2\pi\,z^{|n|}}{\sqrt{a^2-1}},
\qquad z=a-\sqrt{a^2-1},\qquad a>1 ,
\label{eq:poissonkernel}
\end{equation}
at $a=\cosh(\delta/2)$, where $\sqrt{a^2-1}=\sinh(|\delta|/2)$ and $z=e^{-|\delta|/2}$. Expanding $1/(1-e^{-|\delta|})$ gives \eqref{eq:pin}, with one two-sided exponential per collinear pole.

The radial transform of one layer is
\begin{equation}
\int_{-\infty}^{\infty}\big(1-\cos\nu\delta\big)\,e^{-r|\delta|}\,d\delta
=\frac{2}{r}-\frac{2r}{r^2+\nu^2}
=\frac{2\nu^2}{r\,(r^2+\nu^2)} ,
\label{eq:layertransform}
\end{equation}
and summing over $m$ with $r=r_{m,n}$ and
$\psi(z)=-\gamma_E+\sum_{m\ge0}\big[(m+1)^{-1}-(m+z)^{-1}\big]$
gives $\psi(a_n+i\nu)+\psi(a_n-i\nu)-2\psi(a_n)$. The gap between the $n$ and $n=0$ sectors at zero frequency gives the remaining constant,
\begin{equation}
\int_{-\infty}^{\infty}\big[\pi_0(\delta)-\pi_n(\delta)\big]\,d\delta
=\sum_{m\ge0}\Big[\frac{2}{m+\tfrac12}-\frac{2}{m+a_n}\Big]
=2\big[\psi(a_n)-\psi(\tfrac12)\big]=\kappa_n ,
\label{eq:kappaderiv}
\end{equation}
a difference that converges although the separate integrals diverge, and replaces $2\psi(a_n)$ by $2\psi(\tfrac12)$.

Negative definiteness of the exponent follows from the same layers, now with their joint angular and radial dependence. These layers, $\rho=\sum_{m\ge0}\rho_m$ with $\rho_m(\delta,\phi)=e^{-(m+\frac12)|\delta|}\,\rho(\delta,\phi)/\pi_0(\delta)$, are non-negative and each has finite intensity $2/(m+\tfrac12)$. Each $\rho_m$ generates a compound Poisson semigroup with a negative-definite exponent on $\hat{\mathbb G}$. The exponent $q$ is the sum of these layer exponents and is therefore negative definite. A direct proof of negative definiteness of $q$ uses
\begin{equation}
\sum_{j,k}c_j\bar c_k\big(1-e^{i\langle\xi_j-\xi_k,u\rangle}\big)
=-\Big|\sum_jc_je^{i\langle\xi_j,u\rangle}\Big|^2\le0
\qquad\text{whenever}\quad \sum_jc_j=0 ,
\label{eq:negdef}
\end{equation}
whose sign is preserved by integration against the non-negative $\rho$. The Schoenberg correspondence gives the convolution semigroup, and the Hunt theorem identifies its generator on the abelian Lie group $\mathbb G$.

\section{Connections of the L\'evy measure and finite variation}
\label{sec:supp-measure}

Marchesini and Onofri \cite{MarchesiniOnofri} combine conjugation and growth removal at zero conformal spin in their Eq.~(3.1),
$\phi_0(x,\tau)=e^{x/2}e^{-4\ln2\,\tau}T(b_\perp^2,\tau)$.
Here $b_\perp$ is their impact parameter and $b_\perp^2=e^{-x}$. The factor $e^{x/2}=1/b_\perp$ conjugates the amplitude, while $e^{-4\ln2\,\tau}$ removes the growth at the intercept, $4\ln2$ per unit of their variable $\tau=\bar\alpha Y$. Their Eq.~(3.2) gives the resulting gain-loss equation, whose kernel $1/(2\sinh\tfrac12|y-x|)$ is \eqref{eq:pin} at $n=0$.

Del Duca and Schmidt \cite{DelDucaSchmidt2} give the real-emission function of Section~\ref{sec:generator} in their Eq.~(9). The function multiplies the pure functions in Eq.~(4.10) of Del Duca, Dixon, Duhr and Pennington \cite{DDDP}, the first of which is one in their Eq.~(4.12). The function $|w|/|1+w|^2$ equals $2\pi\rho$ in the conventions of Ref.~\cite{DDDP} upon identifying $|w|$ with $k/k'$ and $\arg w$ with $\phi-\pi$.

The step measure can also be identified in two other bodies of work, on known families of L\'evy processes and on a conformal operator on the cylinder. Its radial projections are members of the beta family of Kuznetsov \cite{Kuznetsov}, with unit scale, unit family parameter and decay rate $a_n$. The unit value of the family parameter is admissible. It is also one of the two values where the individual Euler Beta functions in Kuznetsov's characteristic exponent diverge but their difference stays finite. The limit is a difference of digammas, which explains why \eqref{eq:q} has that form instead of a ratio of Gamma functions. On one half-line, the radial law is the subordinator of M\"ohle \cite{Mohle} at parameter $a_n$, evolved at speed $1/a_n$. It is the homogeneous case of the log-beta process of Walker and Muliere \cite{WalkerMuliere}.

The conformal fractional Laplacian gives another identification. DelaTorre, del Pino, Gonz\'alez and Wei \cite{DelaTorreDelPinoGonzalezWei} express this operator on the cylinder as a principal-value integral with an explicit kernel integrated over the angular variables, and show that this kernel is strictly positive and has tempered-stable asymptotics at fixed order in $(0,1)$. Their kernel equals $2\pi$ times the azimuthal integral of \eqref{eq:rho} in two dimensions and at order zero, once their axial coordinate is set to half of $\delta$. Ao, DelaTorre and Gonz\'alez \cite{AoDelaTorreGonzalez} retain the angular dependence and give the unintegrated kernel $2\pi\rho(\delta,\phi)$.

The exponent \eqref{eq:q} follows by differentiating the symbol with respect to the operator order at zero. DelaTorre and Gonz\'alez \cite{DelaTorreGonzalez} give that symbol in Theorem 1.1 as a ratio of squared Gamma moduli, depending on the harmonic index and a Fourier variable twice $\nu$. In two dimensions its order derivative at zero is $2\ln2+2\psi(1)-\chi(n,\tfrac12+i\nu)$. Subtracting the value at the origin removes the constant and gives \eqref{eq:q}. This derivative defines the conformal logarithmic Laplacian, whose sphere spectrum is given by Fern\'andez and Salda\~na \cite{FernandezSaldana}. The identifications relate the measure to laws with known fluctuation theory, although the results of Sections~\ref{sec:generator} to~\ref{sec:removal} do not depend on these connections.

Finite variation follows from the explicit small-step bound
\begin{equation}
\int_{|u|<\eta}|u|\,\rho\,d\delta\,d\phi
\le\frac{2\eta}{1-\eta^2/12}<\infty ,\qquad 0<\eta<2\sqrt3 ,
\label{eq:finitevariation}
\end{equation}
which permits omission of the truncation function, with its contribution absorbed into a drift that vanishes.

Giovannini and Van Hove \cite{GiovanniniVanHove,GiovanniniVanHoveClan} use infinite divisibility in their clan description of the negative binomial multiplicity distribution, a compound Poisson law with a Poisson number of clans and a logarithmic distribution of particles in each clan.

\section{Regularity and iterative solutions}
\label{sec:supp-regularity}

The existence of a density and its boundedness require different arguments. The transform of $\mu_t$ decays at large dual argument because the exponent grows logarithmically. The decay of $\hat\mu_t$ as $|\xi|^{-2t}$ is governed by $t=\bar\alpha Y$. Equation \eqref{eq:uniform} below implies that $\mu_t$ has a square-integrable density for $t>\tfrac12$ and a bounded continuous density for $t>1$, or $Y>1/(2\bar\alpha)$ and $Y>1/\bar\alpha$.

A density exists even below the first threshold. Cutting $\Pi$ in \eqref{eq:triple} at $\eta_\delta$ splits $\mu_t$ into events with at least one step above $\eta_\delta$ and events with none. The first contribution is absolutely continuous, since its resolved-step measure is. The second has weight $e^{-t\Pi(|\delta|>\eta_\delta)}$. This weight vanishes as $\eta_\delta\to0$ by \eqref{eq:activity}, leaving no singular contribution.

Below the boundedness threshold, the density diverges at coincident momenta. A transform falling as $|\xi|^{-2t}$ in two dimensions gives the behavior $|u|^{2t-2}$. This power reaches the boundedness threshold at $t=1$. As $t\to0$, the power tends to $|u|^{-2}$, the behavior of \eqref{eq:nearid}. The law approaches the identity, while its density $p_t$ away from the identity is $t\rho+o(t)$. The density is bounded when $t$ exceeds $1$, with a logarithmic divergence at $t=1$, so boundedness requires enough rapidity for the walk to spread. The coincidence singularity is integrable at every positive rapidity, so no probability is concentrated at a point.

The large-argument expansion of $q$ that determines these properties of $\mu_t$ is uniform in both dual variables,
\begin{equation}
q(\nu,n)=\ln\Big(\nu^2+\frac{n^2}{4}\Big)-2\psi(\tfrac12)+O\big(|\xi|^{-2}\big),
\label{eq:uniform}
\end{equation}
so $\hat\mu_t$ decays as $|\xi|^{-2t}$. Square integrability gives an $L^2$ density for $t>\tfrac12$, and integrability gives a bounded continuous density for $t>1$.

The L\'evy-It\^o theorem makes the counterpart of the factorization used by iterative solutions of \eqref{eq:bfkl} an identity for the process \eqref{eq:triple}. Schmidt \cite{Schmidt}, Andersen and Sabio Vera \cite{AndersenSabioVera}, and Chachamis and Sabio Vera \cite{ChachamisSabioVera} separate emissions resolved above a small parameter from an exponentiated Regge form factor that accounts for the rest. A cut $|\delta|>\eta_\delta$ separates a compound Poisson process with rate \eqref{eq:activity} from an independent remainder of infinitely many shorter steps. The form factor is the probability of no resolved step. Both factors depend on the cut, while their convolution is independent of it.

The resolved-step decomposition has the same shape as the decomposition of iterative BFKL solutions without being the same decomposition. Their usual resolution parameter is an absolute transverse scale, while the cut here is on a transverse-momentum ratio. Schmidt \cite{Schmidt} obtains a Poisson law for a modified equation that replaces each accumulated momentum by the external momentum and bounds every transverse integration above by the square of that external momentum. A cut on the layer index of \eqref{eq:pin}, in place of the cut on the radial displacement, gives a further decomposition. The remainder has the same form with a shifted index, since $\sum_{m>M}e^{-(m+a_n)|\delta|}=\pi_{n'}(\delta)$ with $a_{n'}=a_n+M+1$.

\section{Locations of the fixed-order failures and moment thresholds}
\label{sec:supp-obstruction}

The first failure of convexity occurs further from the collinear edge than that of the weaker strip-edge inequality. Let $\epsilon$ be the distance to the nearer edge, equal to $1-\gamma$ near $\gamma=1$ and to $\gamma$ near $\gamma=0$. At $\bar\alpha=\scanabarA$, convexity \eqref{eq:convexity} first fails at $\epsilon=\convfaileps$, compared with $\stripfaileps$ for \eqref{eq:stripineq}. The next-to-leading eigenvalue terms there are respectively $\convfailratio$ and $\stripfailratio$ times the leading terms. As the coupling decreases, the convexity failure approaches $\epsilon=\sqrt{3\bar\alpha}$ and the strip-edge failure approaches $\epsilon=\sqrt{\bar\alpha/2}$. Their next-to-leading to leading ratios tend to $-\tfrac16$ and $-1$.

The square-root resummation $\sqrt{\epsilon^2+2\bar\alpha}-\epsilon$ of Salam \cite{Salam} expands in $\bar\alpha/\epsilon^2$. It reproduces the unit-residue pole of $\chi_0$ and the cubic coefficient $d_{13}$, with convergence radius $\bar\alpha/\epsilon^2=\tfrac12$ set by its branch point. The parameter equals $2$ at $\epsilon=\sqrt{\bar\alpha/2}$ and $\tfrac13$ at $\epsilon=\sqrt{3\bar\alpha}$. Thus, at small coupling, the first failure of \eqref{eq:stripineq} lies outside the convergence radius and the first failure of \eqref{eq:convexity} lies inside it. Positivity is therefore lost while the expansion reproducing $d_{13}$ still converges.

Both failures at $\bar\alpha=\scanabarA$, just below $\bar\alpha_c$, lie closer to the symmetric point than these limits predict. Their eigenvalue ratios therefore differ from the limiting values. The all-poles sum \eqref{eq:allpoles} in Section~\ref{sec:removal} uses the same square-root function, and its inverse density still changes sign.

The moment identity \eqref{eq:momentid} rests on differentiating the exponent under the integral over the L\'evy measure. Holomorphy of $\chi$ implies that the law at fixed rapidity would have finite exponential moments on the strip \cite{Lukacs}. Sato \cite{Sato} gives the equivalence, for an infinitely divisible law, between this finiteness and finiteness of the same exponential integral against the L\'evy measure away from the origin. Near the origin, the factor $\delta^2$ makes the moment integrals convergent for any L\'evy measure. Together these facts justify this differentiation. The transfer of exponential integrability to the L\'evy measure is needed for \eqref{eq:momentid} at $j\ge2$ and for the tilted measure of \eqref{eq:esscher}.

A bounded correction shows why the growth of the truncated term matters. Truncating the exponential with exponent $-\alpha
e^{-x}$ at first order gives $1-\alpha e^{-x}$, while $\pi_0e^{\pm\alpha e^{-x}}$ are L\'evy densities for every $\alpha$. The truncated factor stays positive for every $x$ up to $\alpha=1$ and becomes negative on $0\le x<\ln\alpha$ above that value. The bounded correction thus has a finite threshold in the parameter, whereas the unbounded correction $-\alpha
x^2/4$ has none.

The Gaussian-damped control density of Section~\ref{sec:obstruction} has even moments
\begin{equation}
M_{2j}(\alpha)=\int_{-\infty}^{\infty}\delta^{2j}\,\pi_0(\delta)\,
e^{-\alpha\delta^2/4}\,d\delta ,\qquad
M_{2j}(0)=\int_{-\infty}^{\infty}\delta^{2j}\,\pi_0(\delta)\,d\delta ,
\label{eq:ctrlmoments}
\end{equation}
and at first order these become $M_{2j}(0)-\tfrac{\alpha}{4}M_{2j+2}(0)$ and cross zero at
\begin{equation}
\alpha_{2j}=\frac{4M_{2j}(0)}{M_{2j+2}(0)} ,
\label{eq:ctrlthresh}
\end{equation}
with thresholds that tend to zero because the moments of an exponentially decaying density grow factorially.

The collinear poles determine the corresponding sequence for \eqref{eq:psinlo}. Removing the polar parts at $\gamma=0$ and $\gamma=1$ leaves both $\chi_0$ and $\chi_1$ holomorphic in $|\gamma-\tfrac12|<\tfrac32$. Expanding the poles about the symmetric point gives
\begin{equation}
\bar\alpha_{2j}=\frac{1}{(2j+1)(2j+2)}\Big[1+\frac{d_{12}}{j+1}+O(j^{-2})\Big]
\ \longrightarrow\ \frac{1}{4j^2} .
\label{eq:thresholdlaw}
\end{equation}
The poles lie a distance $\tfrac12$ from that point, whereas the nearest singularity of the regular part lies a distance $\tfrac32$ away. For every $j\ge1$, the regular contribution to the $2j$-th derivative is suppressed by $(1/3)^{2j}$ relative to the polar contribution. Its prefactor is below $\thrsuppconst$ for $\chi_1$ at $N_c=3$ and $n_f=4$. Within the polar contribution, the cubic term has an extra factor $2j+2$ relative to the double pole and $2(2j+1)(2j+2)$ relative to the simple pole. It outweighs the other poles and regular part together already at $j=1$. The sign at every order is therefore set by $d_{13}$, with $N_c$ and $n_f$ entering the correction.

Contour integration of the even eigenvalue derivatives, including their regular parts, gives the first thresholds $\throne$, $\thrtwo$ and $\thrthree$. The first is $\bar\alpha_c$ in \eqref{eq:abarc}. The sequence has zero infimum over $j$, so the moment identity alone excludes a non-negative density at any fixed positive coupling. It reaches the conclusion of \eqref{eq:stripineq} from derivatives at the symmetric point. The decreasing thresholds nevertheless do not distinguish the kernel from the control density, since in both cases they reflect first-order truncation and the growth of the moments.

The radial completion \eqref{eq:completion} behaves as $x^{-1-2\bar b\bar\alpha}$ at the origin, because its simple-pole coefficients grow at large layer index as $K_{1j}=2\bar b\,H_j+O(1)$, with harmonic number $H_j$. This origin behavior gives the bounds on $\bar\alpha$ that Section~\ref{sec:obstruction} states for integrability against $\min(1,x^2)$, which makes the completion a L\'evy density, and for finite variation. At fixed displacement, $e^{-(j+\frac12)x}$ decays geometrically in the layer index, overcoming the logarithmic growth of $K_{1j}$ and bounded $K_{2j}$, so the sum over layers converges for every $x>0$.

\section{The remainder of the inverted density}
\label{app:remainder}

The first two layer families are kept explicitly, and a contour estimate bounds the remaining layers. Combining the two families gives
\begin{equation}
\pi_{\rm NLO}(x)=e^{-x/2}P_1(x)+e^{-3x/2}P_3(x)+R(x),
\qquad
P_1(x)=\bar\alpha+\bar\alpha^2\Big(\frac{d_{13}}{2}x^2+d_{12}x+d_{11}\Big) .
\label{eq:assembled}
\end{equation}
The polynomial $P_3$ uses the Laurent coefficients at $\gamma=2$ in the same way as the first polynomial, and $R$ denotes the remainder below. Figure~\ref{fig:nlo} shows the crossing of the full density. The polynomial $P_1$ describes only its leading layer.

The inverse transform can be written in terms of the second eigenvalue derivative,
\begin{equation}
\pi(x)=\frac{1}{2\pi i\,x^2}\int_{{\rm Re}\,\gamma=1/2}
\chi''(\gamma)\,e^{(\gamma-\frac12)x}\,d\gamma .
\label{eq:inversion}
\end{equation}
Moving the contour to $\mathrm{Re}\,\gamma=-\tfrac32$ crosses the poles at $\gamma=0,-1$. The remaining contribution obeys
\begin{equation}
|R(x)|\ \le\ \bar\alpha\,\frac{e^{-5x/2}}{1-e^{-x}}
\ +\ \frac{M}{2\pi}\,\bar\alpha^2\,\frac{e^{-2x}}{x^2},
\qquad
M=\int_{-\infty}^{\infty}\big|\chi_1''(-\tfrac32+i\tau)\big|\,d\tau ,
\label{eq:remainder}
\end{equation}
where the first term sums the leading-order tail. The second bounds the next-to-leading contour integral. The constant $M$ is finite because $\chi_1''$ decreases as $\ln\tau/\tau^2$ on this line. Both this decay and the pole-pair decomposition follow by expressing $\chi_1''$ as its principal parts at $\gamma=-j$ and $\gamma=1+j$, for $j\ge0$.

The principal parts leave no entire remainder. The eigenvalue $\chi_1$ has third-order poles at all integers and grows at most as $\ln^2|\gamma|$ away from them. The difference between $\chi_1''$ and the principal-part sum is entire, with growth bounded by $\ln^2|\gamma|$ on $|\gamma|=N+\frac12$, and must be constant. Both terms tend to zero on $\mathrm{Re}\,\gamma=\frac12$, so this constant vanishes.

We use an upper estimate of $M/2\pi$, rounded upward to one decimal place and printed in the archive as $\Motwopiup$. It combines quadrature in a bounded $\tau$ window with a closed-form majorant for the tail. Quadrature over the whole line, tail included, gives the smaller ratio $\Motwopi$. For the window contribution we take the result at the higher of two quadrature degrees and add the difference between the two results. This allowance measures convergence but does not enclose the quadrature error.

The rounded estimate gives, at each coupling, an estimated positivity interval in $x$. On this interval, the full leading-order sum, together with the first two next-to-leading layers in \eqref{eq:assembled}, exceeds the second term of \eqref{eq:remainder}. The leading-order sum already contains the first term of \eqref{eq:remainder}, so it is counted only once. Positivity on the estimated interval holds if the estimate bounds the true ratio. The estimated interval ends before the crossing in Figure~\ref{fig:nlo} at all four couplings, with gaps $\posgapA$, $\posgapB$, $\posgapC$ and $\posgapD$. Every result in Section~\ref{sec:obstruction} is independent of the value of $M$.

\section{Conjugations, resummed densities and asymmetric schemes}
\label{sec:supp-resummation}

The family $g=k^{s}f$ of Section~\ref{sec:removal} contains every continuous pointwise multiplicative conjugation that keeps the kernel translation invariant and the density real and non-negative. Suppose $g=cf$ leaves the kernel translation invariant. Then $c(z+u)/c(z)$ is independent of $z$ for every $u$, making $c$, up to a constant, a continuous homomorphism from $\mathbb G$ to the nonzero complex numbers. Its radial part is $e^{s\ell/2}$ with complex $s$, and its angular part is $e^{in_0\theta}$ with integer $n_0$. A real non-negative density requires $n_0=0$ and real $s$, leaving $g=k^sf$. Continuity is assumed because the functional equation by itself admits nonmeasurable solutions.

The two matching constants of Sabio Vera \cite{SabioVera} in Section~\ref{sec:removal} distinguish the pure and matched forms. Keeping them explicit gives
\begin{equation}
\begin{gathered}
\omega_{A,B}(\gamma)=\sum_{m\ge0}\Big[F_{A,B}(\gamma+m)+F_{A,B}(1-\gamma+m)
-\frac{2\bar\alpha(1+A\bar\alpha)}{m+1}\Big],\\[2pt]
F_{A,B}(\lambda)=\sqrt{\lambda_B^2+2\bar\alpha(1+A\bar\alpha)}-\lambda_B ,
\qquad \lambda_B=\lambda+B\bar\alpha .
\end{gathered}
\label{eq:allpolesAB}
\end{equation}
At $A=B=0$ it reduces to \eqref{eq:allpoles}, the approximation to the leading-logarithmic shift. Sabio Vera \cite{SabioVera} quotes its intercept at $\bar\alpha=0.2$ as $0.35$. Next-to-leading matching instead sets $A=d_{11}$ and $B=-d_{12}$ from \eqref{eq:dvalues}, giving $\omega_{\rm m}=\omega_{d_{11},-d_{12}}$. The layer decay rate becomes $m+\tfrac12+B\bar\alpha$ and the Bessel argument becomes $a_{\rm m}=\sqrt{2\bar\alpha(1+d_{11}\bar\alpha)}$. The matched density is
\begin{equation}
\pi^{\rm m}_{\rm eff}(x)=\frac{e^{-(\frac12+B\bar\alpha)x}}{1-e^{-x}}\,
\frac{a_{\rm m}J_1(a_{\rm m}x)}{x},\qquad B=-d_{12} .
\label{eq:allpolesmatched}
\end{equation}
The positive factor $e^{-B\bar\alpha x}$ changes the damping but leaves the zeros fixed. The density changes sign for every $0<\bar\alpha<1/|d_{11}|$. At $\bar\alpha=\scanabarD$, its first zero moves from $\bessconst/\sqrt{\bar\alpha}$ to $\zeromatchhi$, solely through $d_{11}$ in $a_{\rm m}$.

The bound $1/|d_{11}|$ depends on flavor number because $d_{11}$ is proportional to $n_f$, and lies far outside the physical range for every $n_f\ne0$. There is no finite bound at $n_f=0$, where matching moves no zero, although $d_{12}=-11/8$ still produces damping. Above $1/|d_{11}|$, imaginary $a_{\rm m}$ turns the Bessel factor into $-|a_{\rm m}|I_1(|a_{\rm m}|x)/x$, negative for every $x$. Positivity then fails without oscillation. At $\bar\alpha=1/|d_{11}|$, the factor $1+d_{11}\bar\alpha$ vanishes in both the density and the virtual subtraction. The exponent $\omega_{\rm m}$ vanishes throughout the strip, so at this one coupling the step measure is the zero measure, non-negative only because it vanishes, and the process never jumps.

The size of the negative contributions to the pure and matched all-poles densities can be measured by their share of radial displacement. An unweighted integral would depend on the lower cut because $\pi_{\rm eff}\sim\bar\alpha/x$ at the origin. The weighted integral $2\int_0^\infty x\,|\pi_{\rm eff}(x)|\,dx$ converges at both ends. It gives the mean accumulated radial displacement per unit rapidity for a non-negative density. Before matching, the fraction contributed by the negative region is $\lobevarhi$ at $\bar\alpha=\scanabarD$ and $\lobevarlo$ at $\bar\alpha=\scanabarA$. Matching changes these to $\lobematchhi$ and $\lobematchlo$.

The factor $e^{-x/2}$ suppresses the negative fraction at a first zero that recedes as $\bar\alpha^{-1/2}$. The negative fraction therefore decreases exponentially as the coupling falls, while remaining nonzero. The first unmatched zero corresponds to $k'/k=\ratiohi$ at $\bar\alpha=\scanabarD$, with damping $\damphi$, and to $\ratiolo$ at $\bar\alpha=\scanabarA$, with damping $\damplo$. Matching moves the ratio at $\bar\alpha=\scanabarD$ to $\ratiomatchhi$.

The full prescription of Sabio Vera \cite{SabioVera}, tested in Section~\ref{sec:removal}, includes $\bar\alpha\chi_0+\bar\alpha^2\chi_1$ and subtracts its terms already present in the resummed expression. Summing with the pole representation of $\chi_0$ reduces the result to
\begin{equation}
\omega_{\rm AP}=\omega_{\rm m}+\bar\alpha^2\tilde\chi_1,\qquad
\tilde\chi_1=\chi_1-d_{11}\chi_0-d_{12}\big[\psi'(\gamma)+\psi'(1-\gamma)\big]
-\tfrac14\big[\psi''(\gamma)+\psi''(1-\gamma)\big],
\label{eq:allpolesclosed}
\end{equation}
where $\omega_{\rm m}$ is the matched expression \eqref{eq:allpolesAB}. The three subtractions remove the simple, double and cubic collinear poles of $\chi_1$, respectively. The remainder $\tilde\chi_1$ is regular at $\gamma=0$ and $\gamma=1$, with nearest poles of double order at $\gamma=-1$ and $\gamma=2$. The pole at $\gamma=2$ follows from that at $\gamma=-1$ by evenness of $\chi$ under $\gamma\to1-\gamma$.

This prescription removes the cubic pole in every layer, because the leading Laurent coefficient of $\chi_1$ is $d_{13}$ at each $\gamma=1+j$, including $\gamma=1$. Each remaining double-pole layer gives $(c_{1j}+c_{2j}x)\,e^{-(j+\frac12)x}$, with large-$x$ sign set by $c_{2j}$. Both coefficients are positive above the first layer, while regularity of $\tilde\chi_1$ at the first layer makes its contribution zero. The correction is positive for every $x>0$. Its coefficients are tabulated for sixty-four layers in the archive. Their closed form gives $c_{2j}\ge\clayertwoinf$ and $c_{1j}\ge\clayeronebound$ in every layer above the first.

The full density still changes sign at both computed couplings. Its negative fractions are $\lobeaphi$ at $\bar\alpha=\scanabarD$ and $\lobeaplo$ at $\bar\alpha=\scanabarA$, with first zeros $\zeroaphi$ and $\zeroaplo$. Both zeros lie beyond the pure all-poles value $\bessconst/\sqrt{\bar\alpha}$. The $\bar\alpha^{-1/2}$ scaling is recovered as the coupling decreases.

The kernel of Iancu et al.\ \cite{IMMST} in Section~\ref{sec:removal} has no anticollinear partner or virtual subtraction. Their full shift by $\omega$ accounts for the factor of two in the coupling, relative to the symmetric shift $\omega/2$. They credit the kernel's earlier appearance to Sabio Vera \cite{SabioVera}. The kernel's first sign change is closer to the origin than $x^{\rm B}_*$ of Section~\ref{sec:removal} by a factor $1/\sqrt2$. The interval in which complete monotonicity fails is a factor $\sqrt2$ wider than for \eqref{eq:allpoles}. Their local formulation also requires a modified initial condition that oscillates along with the kernel.

The gluon-quark mixing rates in the resummation of Kovner, Lublinsky, Nefedov and Skokov \cite{KLNSone} give the nonzero-flavor version of the positive-density test of Section~\ref{sec:removal}. The two rates for $n_f\ne0$ are $-\alpha_s/(4\pi)$ times the gluon-quark mixing eigenvalues of that resummation, and one of them has vanishing weight at zero flavors. The resulting two-exponential condition holds at all four tested couplings for $n_f=3$ and $n_f=4$. The failure onset rises with flavor number to $\acabarfour$ at four flavors. The pole offsets sum to $a_s(11N_c+2n_f)/3$, where $a_s=\alpha_s/(4\pi)$ in Ref.~\cite{KLNSone}, while \eqref{eq:conventions} uses $\alpha_s$. Kovner, Lublinsky, Nefedov and Skokov \cite{KLNStwo} treat running-coupling effects in this resummation separately.

The asymmetric-scheme failures follow from the poles left by the scale change in Section~\ref{sec:removal}. At small coupling, convexity first fails near the edge at which the cubic pole cancels, at $\epsilon=-3\bar\alpha d_{12}$, where the next-to-leading term is $-\tfrac13$ of the leading term. At the opposite edge, $\gamma\to0$, the shift doubles $d_{13}$. The onset then approaches $\epsilon=\sqrt{6\bar\alpha}$, compared with $\sqrt{3\bar\alpha}$ in the symmetric scheme, and the ratio tends to $-\tfrac16$.

The shift leaves $\chi_1''(\tfrac12)$ unchanged because $\chi_0'(\tfrac12)$ and $\chi_0'''(\tfrac12)$ vanish. Thus $\bar\alpha_c$ in \eqref{eq:abarc} is the same in the two schemes. All three scanned couplings above $\bar\alpha_c$ fail at the symmetric point in both. At $\bar\alpha=\scanabarA$, the first failures are at $\epsilon=\plusepshiA$ from $\gamma=1$ and $\epsilon=\plusepsloA$ from $\gamma=0$, compared with $\convfaileps$ at either symmetric-scheme edge. Only convexity \eqref{eq:convexity} applies, because $\chi$ is not even in the $(+)$ scheme.

At $\bar\alpha=\plusabarD$, the onsets are $\epsilon=\plusepshiD$ and $\epsilon=\plusepsloD$, compared with the limiting values $-3\bar\alpha d_{12}$ and $\sqrt{6\bar\alpha}$. The respective ratios are $\plusratiohiD$ and $\plusratioloD$. The lower-sign choice gives $\chi_1^{(-)}=\chi_1-\tfrac12\chi_0\chi_0'$. Evenness of $\chi_0$ and $\chi_1$ makes $\chi_0\chi_0'$ odd, so $\chi_1^{(-)}(\gamma)=\chi_1^{(+)}(1-\gamma)$. The $(-)$-scheme failures are the reflections of these values.

The fixed-order convexity failure extends to an energy scale $s_0=q_1q_2$ multiplied by any real power of $q_1/q_2$. Powers $\pm1$ give the two asymmetric shifts. At fixed $\omega$, the amplitude is multiplied by this momentum ratio to a power proportional to $\omega$. This multiplication translates the Mellin variable conjugate to the logarithm of the squared ratio. Substitution of $\omega=\bar\alpha\chi_0$ changes $\chi_1$ by a multiple of $\chi_0\chi_0'$, with coefficient half the scale power up to the Mellin orientation sign. The two cubic coefficients shift equally and oppositely. Each equals $d_{13}$ in the symmetric scheme, so their sum remains $2d_{13}$ for every power and for both orientation signs. Because $d_{13}<0$, at least one stays negative. The curvature $\chi''$ tends to $-\infty$ at that edge, with the shifted coefficient replacing $d_{13}$ in the Laurent argument of Section~\ref{sec:obstruction}. Convexity fails at fixed order for every scale in this family at any fixed positive coupling.

A larger state space cannot give a positive law with the same cylinder projection as the fixed-order law excluded in Section~\ref{sec:obstruction}, since a measurable image of a non-negative measure remains non-negative and that law would have characteristic function $e^{-Y\Psi}$. Rapidity dependence of translation-invariant L\'evy generators cannot evade the exclusions on the cylinder either, since integrating their negative-definite exponents over rapidity preserves negative definiteness and hence infinite divisibility, which is the property that fails.

\section{Finite-rapidity tests and the contour assumption}
\label{sec:supp-finite}

We test the collinearly improved Green functions of Colferai, Li and Sta\'sto \cite{ColferaiLiStasto} at $N_c=3$ and $n_f=4$, four couplings $\bar\alpha=\scanabarA$, $\scanabarB$, $\scanabarC$ and $\scanabarD$, and five times $t=\bar\alpha Y=\scantA$, $\scantB$, $\scantC$, $\scantD$ and $\scantE$. The frequency spacing is $\Delta\nu=\rgiDeltanu$. One grid contains $\nu=0,\Delta\nu,\dots,4\Delta\nu$, and the other has twenty-one frequencies through $\nu=5$. The positivity conclusions in the archive are stated for the twenty-one-frequency grid. The couplings in physical units are $\alpha_s=\bar\alpha\pi/N_c=\alphasScanA$, $\alphasScanB$, $\alphasScanC$ and $\alphasScanD$.

The shortest time gives negative smallest eigenvalues on the five-point grid for all three prescriptions and all four couplings, subject to the contour assumption of Section~\ref{sec:removal}, which the root count below supports. Their values are
\begin{equation}
\begin{array}{lcccc}
\bar\alpha & \scanabarA & \scanabarB & \scanabarC & \scanabarD\\[2pt]
\mathrm{collA} & \rgimeAaA & \rgimeAbA & \rgimeAcA & \rgimeAdA\\
\mathrm{collB} & \rgimeBaA & \rgimeBbA & \rgimeBcA & \rgimeBdA\\
\mathrm{zV} & \rgimeZaA & \rgimeZbA & \rgimeZcA & \rgimeZdA
\end{array}
\label{eq:bochgrid}
\end{equation}
at $t=\scantA$. Every violation at the two larger couplings exceeds the smallest by more than two orders of magnitude. Section~\ref{app:numerics} gives a coefficient vector realizing the smallest violation and estimates its error. The corresponding rapidities are $Y=t/\bar\alpha=\rgiYa$, $\rgiYb$, $\rgiYc$ and $\rgiYd$. Holding $t$ fixed therefore varies the rapidity with the coupling. The strongest violations occur at the shorter rapidities $\rgiYc$ and $\rgiYd$. We give greater weight to the tests at $\rgiYa$ and $\rgiYb$, where the violations are milder but the rapidities are larger.

The persistence of a negative smallest eigenvalue depends on the frequency grid. The five-frequency matrix gives $\cellsA$, $\cellsB$, $\cellsC$, $\cellsD$ and $\cellsE$ negative minima among the twelve prescription-and-coupling combinations as time increases through the five values listed above. This matrix is a principal submatrix of the twenty-one-frequency matrix, so every cell negative on the smaller grid is negative on the larger. The larger-grid counts agree for the first three times and become $\cellswideD$ and $\cellswideE$ for the last two.

The longest-time violations are collB and zV at $\bar\alpha=\scanabarD$. Both are negative at every tested time and end at $\rgiwideBdE$ and $\rgiwideZdE$. Their estimated error $\rgieigboundwide$ is formed at the smallest coupling, where $|P_t|\le1$, and at the shortest time. The normalization at the two longest-time points is more than an order of magnitude larger, making this estimate an overestimate there. The larger-grid test leaves $\cellswidetotal$ of the sixty prescription, coupling and time combinations unresolved, compared with $\cellstotal$ on the smaller grid.

The leading-order exponent, whose L\'evy measure is non-negative, gives the smallest five-point eigenvalue $\ablstageone$ at $\bar\alpha=\scanabarA$ and $t=\scantA$. Including successive terms at the same coupling and time locates the failure. The shift of the digammas by $\omega/2$ at $U=0$ leaves $\ablstagetwo$ positive. Including $\omega U(\omega)$ changes it to $\ablstagethreeA$, $\ablstagethreeB$ and $\ablstagethreeZ$ in the three prescriptions, before $\bar\alpha X_1$ enters. The coefficient $d_{13}$ comes through the shift, and $d_{12}$ in the zV prescription comes through $\omega U(\omega)$. These stages omit $\bar\alpha X_1$, so their negative eigenvalues arise from the resummed kernel itself. The fixed-order exclusion in Section~\ref{sec:obstruction} instead follows from \eqref{eq:dvalues}, independently of these computed values.

A positive measure also requires $|P_t(\nu)|\le P_t(0)=1$. This weaker bound needs neither a vector nor an eigenvalue calculation, and it is tested on the same twenty-one frequencies. At $t=\scantA$, the grid maximum of $|P_t|$ exceeds unity in $\wideoverone$ of the $\widecells$ cases. The largest value is $\maxPwideAd$, for collA at $\bar\alpha=\scanabarD$ and $\nu=\maxPwideAdnu$. At $\bar\alpha=\scanabarA$, all prescriptions attain their grid maximum at $\nu=0$, where $P_t$ is normalized. The grid detects no failure of this bound at the smallest coupling.

The same inversion and contour assumption test the convolution semigroup identity $P_{t+s}=P_tP_s$. At $t=s=\scantA$ and $\nu=1$, the prescription and coupling with the weakest eigenvalue violation give $P_{t+s}-P_tP_s=\semidefBa$. Over the twelve combinations the differences reach $\semidefAc$. The normalized family $P_t$ therefore fails to form a continuous convolution semigroup.

The contour assumption is supported by an argument-principle count of denominator zeros. In the variable $\lambda=\omega/\bar\alpha$, the denominator of \eqref{eq:rgiGreen} is $\bar\alpha(\lambda-X)$. We track the real zero of $\lambda-X$ reached from the leading-order value $\chi_0(\tfrac12+i\nu)$. The three prescriptions, four couplings and five frequencies of the printed vector give $\rootcountpoints$ points. At each, the winding number of $\lambda-X$ is one on a rectangle around the tracked root and zero on a rectangle starting just to its right.

Both counts stay unchanged as the half-extent grows from $\rootcountreachsmall$ to $\rootcountreach$, and $|\lambda-X|$ stays above $\rootcountmargin$ on each contour. Circles of radius $\rootcountreach$ and larger cover the exterior. The smallest circle radius equals the largest rectangle half-extent, so the two checks leave no gap. We use the circle portions with $\mathrm{Re}\,\lambda\ge\rootcountedge$, to the right of the first pole of $X$ at $-1/\bar\alpha$ for every tested coupling.

Sampling in azimuth and at the limiting abscissa gives $|\bar\alpha X|$ below $\rootcountfar$ times $|\omega|$. Refining the sampling by a factor of sixty-four reproduces this value to thirteen figures. A closed-form bound also controls the large-radius decay. The Binet integral bounds each digamma in $X$, to the right of $\rootcountedge$ and beyond $|\lambda|=\rootcountrzero$, by the logarithm of its argument plus a constant. The resulting bound gives $|X|$ smaller than $|\lambda|$ for every prescription, coupling and frequency used in inversion. The radius of the Binet bound is below $\rootcountreach$, so the bound covers every exterior-circle portion used. The ratio below unity excludes denominator zeros there. The bound and the computed finite-contour counts exclude singularities to the right of the tracked root at the sixty points.

The root must lie to the right of $-1/\bar\alpha$ for these rectangles to contain no kernel pole and hence count zeros alone. It crosses that abscissa above $\nu=2$ at $\bar\alpha=\scanabarD$ and from $\nu=4$ at $\bar\alpha=\scanabarC$. On the wider grid we therefore fix the left edge at $\rootcountedge$. The $\rootcountwidefreq$ frequencies give $\rootcountwidepoints$ points with the same prescriptions and couplings. The count is one if the tracked root lies right of the edge and zero otherwise, unchanged as the half-extent grows from $\rootcountreachsmall$ to $\rootcountreach$.

Across both grids, $|\lambda-X|$ stays above $\rootcountmargin$ on the contours and $|\bar\alpha X|$ stays below $\rootcountfar$ times $|\omega|$ on exterior circles. Thus the tracked root is the only zero to the right of the edge if it lies on that side. All zeros lie left of the edge if the tracked root lies left of it, but their ordering is undetermined. Another zero could then lie to the right of the tracked root, so the count does not identify that root as the rightmost singularity of \eqref{eq:rgiGreen}.

Over the five times, the inversion contour ranges from $\mathrm{Re}\,\lambda=\inversionabscissalo$ to $\inversionabscissahi$, right of both the edge and every counted zero. Every singularity of \eqref{eq:rgiGreen} therefore lies left of the inversion contour in each tested case. The finite-contour counts are computations, and the finite-rapidity conclusions retain the contour assumption of Section~\ref{sec:removal}.

All these tests apply separately to each tested combination of prescription, coupling and time on the stated grids. A negative eigenvalue excludes positive definiteness of $P_t$ for that combination, whereas a non-negative one leaves it unresolved. The largest $|P_t|$ is a discrete maximum, without maximizing between grid points. A semigroup defect alone also does not exclude infinite divisibility of a single $P_t$. The denominator of \eqref{eq:rgiGreen} has no time dependence, so its root count requires no time variable. The exclusions concern the three prescriptions, one family from one paper, at the displayed parameters.

\section{The next-to-leading eigenvalue in the symmetric scheme, the computational protocol and the public archive}
\label{app:numerics}

The authors formulated the physics problem and directed the work at every stage. Generative AI models carried out the technical execution under their direction, including the derivations and calculations, the programs producing the numbers and both figures, independent recomputations and cross-checks, and the drafting of the text. The models were Claude Opus 5, Claude Opus 5.5 and Claude Fable 5.1 from Anthropic, GPT-5.6 Sol and GPT-6 Astra from OpenAI, Gemini 3.1 Pro from Google, and Kimi K3 from Moonshot AI.

The authors specified the computational tasks and checking criteria, ran and inspected the programs, and checked the reported results. The co-authors checked the derivations and numbers. Every number and both figures are reproduced by the programs in the public archive identified in the Data availability statement. The figures were drawn by programs written by these models, and the authors checked them against the computed data. The authors take full responsibility for the content.

Values attributed to prior work, equation numbers in cited papers, scan inputs and figure settings are transcribed. Every other number is computed by the programs or reproduced from the archived results. Eigenvalue evaluation uses arbitrary precision with at least twenty-five working digits, except for the pole-layer reconstruction of $\chi_1$ on the critical line used in finite-rapidity inversion. That reconstruction uses double precision and a fixed layer count. Even derivatives are evaluated by Cauchy integration around $\gamma=\tfrac12$ on circles inside the nearest poles, including one of radius $\cauchyrad$. The leading-order closed form serves as a control.

The finite-rapidity inverse of \eqref{eq:rgiGreen} uses the arbitrary-precision method of de Hoog, Knight and Stokes \cite{deHoog} on the vertical contour described in Section~\ref{sec:removal}. The archive gives its order and the resulting $P_t$ values. The four rapidities in Section~\ref{sec:supp-finite} are recomputed from $t/\bar\alpha$. Comparisons use the fixed contour of Talbot \cite{Talbot} and a second de Hoog order. All three calculations use the same pole-layer reconstruction of $\chi_1$, described above, and agree to the quoted digits at all five times. The difference of this reconstruction from the closed form increases with $\nu$ and is given in the archive.

The weakest of the twelve short-time negative eigenvalues is collB at $\bar\alpha=\scanabarA$, subject to the contour assumption of Section~\ref{sec:removal}. A unit coefficient vector on the five frequencies is
\[
(\rgiwitBzero,\,\rgiwitBone,\,\rgiwitBtwo,\,\rgiwitBthree,\,\rgiwitBfour),
\]
which gives quadratic form $\rgiwitBform$ against the stored $P_t$ values. To estimate the inversion error, we compare the primary de Hoog order with a much higher reference order and take the largest difference over the inversion frequencies and times. A further increase of the reference order changes the result by much less.

The difference affects both numerator and denominator in the normalization of $P_t$. Multiplication by the matrix size, using the Weyl inequality described by Horn and Johnson \cite{HornJohnson}, gives estimated eigenvalue errors $\rgieigbound$ for five frequencies and $\rgieigboundwide$ for twenty-one. Agreement between orders measures convergence but does not enclose the error relative to the inverse transform, so both figures are error estimates and not rigorous bounds. The primary and reference calculations agree in every stored digit at the five frequencies of the displayed vector.

Agreement between methods does not by itself locate the singularities. Two inversions can miss the same singularity and agree while both are wrong. The Talbot contour opens into the left half-plane and must enclose all singularities of \eqref{eq:rgiGreen}. A conjugate zero pair lies outside it at the longest scanned time, but inside it at the shortest time used for the twelve displayed eigenvalues. The eigenvalue-error estimate therefore relies only on the two de Hoog orders.

The implementation corrects four terms in the printed next-to-leading eigenvalue $\chi_1$ of Colferai, Li and Sta\'sto \cite{ColferaiLiStasto}, Eqs.~(2.7) and (2.9) in the article and (7) and (9) in the preprint. All four misprints occur in both versions. The coefficient of $\chi_0^2$ must be $-\bar b/2$, with $\bar b=(11N_c-2n_f)/(12N_c)=\bbarvalue$ at $N_c=3$ and $n_f=4$. The printed $b$ from Eq.~(2.8) contains a factor $C_A/\pi$, where $C_A=N_c$ is the adjoint Casimir. Also, $\xi(3)$ must read $\zeta(3)$, and $\pi^2/(4\sin\pi\gamma)$ must read $\pi^3/(4\sin\pi\gamma)$. The second denominator of $\Phi$ must be $(n+1-\gamma)^2$, the reflection of the first. Its numerator $\psi(n+2-\gamma)$ is already reflected.

Fadin and Lipatov \cite{FadinLipatov} give three corrections in Eq.~(14) and the fourth in Eq.~(15). Their Eq.~(12) fixes the normalization, with their $\delta$ equal to four times $\chi_1$ here. The overall minus sign on the bracket in Eq.~(14) turns $\tfrac12(11/3-2n_f/(3N_c))\chi^2$ into $-\bar b\chi_0^2/2$ after division by four. The terms $-6\zeta_3$ and $-\pi^3/\sin\pi\gamma$ similarly give $\tfrac32\zeta_3$ and $\pi^3/(4\sin\pi\gamma)$. Equation~(15) gives $(n+1-\gamma)^2$ as the second denominator of $\Phi$. An independent transcription of their expression agrees with the implementation to twenty-five digits on a grid of real arguments inside $0<\mathrm{Re}\,\gamma<1$ and is included in the archive. Kotikov and Lipatov \cite{KotikovLipatov} give the eigenvalue in closed form in $\gamma$ and conformal spin $n$. Their $n=0$ expression agrees with Fadin and Lipatov \cite{FadinLipatov}.

The implementation and the independent transcription both use the even part of the eigenvalue, omitting the term $\psi'(\gamma)-\psi'(1-\gamma)$ of Fadin and Lipatov. That term comes from the running-coupling argument. Their Eq.~(1) already uses the symmetric scale $s_0=qq'$, and an asymmetric energy scale gives a different odd shift, $\pm\tfrac12\chi_0\chi_0'$. The eigenfunction redefinition below their Eq.~(15) removes the running-coupling term. Multiplying the basis $q^{2(\gamma-1)}$, with renormalization scale $\mu$, by
$\big(\alpha_s(q^2)/\alpha_s(\mu^2)\big)^{-1/2}=1+\tfrac12\bar b\,\bar\alpha\ln(q^2/\mu^2)$
changes $\chi_1$ to $\chi_1+\tfrac12\bar b\,\chi_0'$. This cancels $-\tfrac12\bar b\,\chi_0'$, since $\chi_0'=-\psi'(\gamma)+\psi'(1-\gamma)$.

The corrected expression agrees with Sabio Vera \cite{SabioVera} and gives $\chionepp$ for $\chi_1''(\tfrac12)$. The $\pi^3$ correction changes this curvature by $\shiftpi$, the correction to the denominator of $\Phi$ changes it by $\shiftphi$, and the correction to the coefficient of $\chi_0^2$ changes it by $\shiftb$. The $\xi$ correction is constant and leaves the curvature unchanged.

The eigenvalue calculation of Section~\ref{sec:removal} covers both grids at all sixty prescription, coupling and time combinations. Each of the twelve displayed five-point violations has an archived coefficient vector with a negative quadratic form, subject to the contour assumption. Its sign can therefore be checked directly from the stored inverse without repeating the inversion.

We integrate the negative parts of the pure and matched all-poles densities between successive Bessel zeros with at least twenty-five working digits. For the full prescription, a sign scan followed by bisection determines the interval boundaries. The shape and the integrated weight of its correction converge at different rates. The first two dozen layers of the correction determine the zeros and negative regions but not the total weight.

The simple-pole coefficients grow as $2\bar b$ times the logarithm of the layer index, solely from $-\tfrac12\bar b\,\chi_0^2$. Thus the tail of $\int_0^\infty x\,D(x)\,dx$, where $D$ is the inverse density of $\tilde\chi_1$, is of order $\ln
N/N$. Sixteen layers give $\weightsixteen$, compared with the limit $\weightlimit$. We sum four hundred layers and include a logarithmic tail whose two coefficients are fitted between layers two hundred and four hundred. The resulting weight is fixed to a part in a thousand, so the negative shares of the full prescription are quoted to three figures.

The public archive consists of the ancillary files of arXiv:2609.18269. Its root list gives a checksum for each file. It contains the programs, stored inputs and generated files for the $\archiveprinted$ quoted numbers, as well as the root-count contours, winding numbers, tolerances and bounds.

The assembly commands in \texttt{anc/CONTENTS.txt} place the programs and manuscript source together in a writable directory. The programs reproduce all quoted numbers and both figures, using stored inversion results for the finite-rapidity values. Sixteen quoted numbers come from three stored files. Two programs that print no macros recompute the four positivity gaps and two values of $M/2\pi$ in Section~\ref{app:remainder}, and the five successive-term values in Section~\ref{sec:supp-finite}. Other archived programs recompute the remaining five under different names, with agreement in each case.

\end{document}